\documentclass[10pt]{article}
\usepackage{amsmath}
\usepackage{graphicx}
\usepackage{amsfonts}
\usepackage{amssymb}
\usepackage{epsf}
\usepackage{latexsym}

\def\be{\begin{equation}}
\def\ee{\end{equation}}
\def\bea{\begin{eqnarray}}
\def\eea{\end{eqnarray}}
\def\pa{\partial}

\def\fn{\footnote}

\def\tr{\mbox{\scriptsize trial}}
\def\d{\textrm{d}}

\def\case#1/#2{\textstyle\frac{#1}{#2}}

\begin{document}

\begin{titlepage}

\vspace{.7in}

\begin{center}
\Large
{\bf ON THE RECOVERY OF GEOMETRODYNAMICS}

\vspace{.15in}

{\bf FROM TWO DIFFERENT SETS OF FIRST PRINCIPLES}
\normalsize

\vspace{.3in}

{\bf Edward Anderson}

\vspace{.3in}

\noindent{\em Peterhouse, Cambridge, U.K., CB21RD;}

\noindent{\em DAMTP, Centre for Mathematical Sciences, Wilberforce
Road, Cambridge, U.K., CB30WA. Email address: ea212@cam.ac.uk}

\noindent This work was started at {\em the Astronomy Unit, School of Mathematical Sciences,
Queen Mary, University of London, U.K., E14NS}, where EA also has had Visitor Status, and 
was completed at {\em Department of Physics, P-412 Avadh Bhatia Physics Laboratory,
University of Alberta, Edmonton, Canada, T6G2J1.}

\end{center}

\normalsize

\vspace{.3in}

\begin{abstract}

The conventional spacetime formulation of general relativity may be recast as a dynamics of
spatial 3-geometries (geometrodynamics).
Furthermore, geometrodynamics can be derived from first principles.
I investigate two distinct sets of these:
(i) Hojman, Kucha\v{r} and Teitelboim's, which presuppose that the spatial
3-geometries are embedded in spacetime.
(ii) The 3-space approach of Barbour, Foster, \'{O} Murchadha and Anderson in which
the spatial 3-geometries are presupposed but spacetime is not.
I consider how the constituent postulates of the conventional approach
to relativity emerge or are to be built into these formulations.
I argue that the 3-space approach is a viable description of classical physics
(fundamental matter fields included),
and one which affords considerable philosophical insight because of its `relationalist' character.
From these assumptions of less structure, it is also interesting that conventional relativity can
be recovered (albeit as one of several options).
However, contrary to speculation in the earlier 3-space approach papers, I also argue that this
approach is not selective over which sorts of fundamental matter physics it admits.
In particular, it does not imply the equivalence principle.

\end{abstract}

\vspace{.3in}


%




\mbox{ }


\end{titlepage}

\section{Introduction}

While Einstein arrived at general relativity (GR) by reconciling the spacetime notion of
special relativity (SR) with the laws of gravitation by passing to curved spacetime,
Wheeler (1968) and Misner, Thorne \& Wheeler (1973) (MTW) consider this to be but
the first of six routes to relativity (and yet more such routes are now known).
This article considers dynamical routes to GR, which are an interesting possibility
because the spacetime paradigm differs from the otherwise dynamical development of modern physics.
The first evidence for dynamical routes to GR lies in its reformulability
as a dynamics of spatial 3-geometries (geometrodynamics).
This article considers the further step of arriving at geometrodynamics from plausible first principles.

I begin by briefly discussing the first (traditional) route to relativity in Sec 2.
This contemporary account (see Einstein 1950, MTW, Stewart, 1991, Rindler, 2001) serves both to define
precisely the terminology of this article for the postulates of conventional relativity and as an
example of how the construction of a route to GR should be carefully studied to reveal the (sometimes
tacit) mathematical simplicities that inevitably accompany the more publicized physical and
philosophical considerations.
In Sec 3 I explain the many-routes position.
In Sec 4 I then explain how GR may be reformulated (Dirac, 1964, Arnowitt, Deser \& Misner, 1962)
as a geometrodynamics (Wheeler, 1962, 1964a, 1968, DeWitt, 1967, MTW, Kucha\v{r} 1976abc, 1977, 1992,
Isham, 1993), and on what insights are gained from this.
I then consider two distinct first-principles derivations of geometrodynamics: the HKT approach
(Teitelboim, 1973ab, Hojman, Kucha\v{r} \& Teitelboim, 1976, Kucha\v{r}, 1974, Teitelboim, 1980)
in Sec 5, and the 3-space approach (TSA) (Barbour, Foster \& \'{O} Murchadha, 2002ab,
Anderson \& Barbour 2002, Anderson, 2003, 2004a, 2005)\fn{I
subsequently refer to all of these as `the TSA papers',
       to the first three of these as `the earlier TSA papers', and
             to the first of these as `the RWR paper' after its title (relativity without relativity).}
in Sec 6.
HKT's first principle is {\sl embeddability into spacetime}, which is implemented by
requiring that prospective theories' constraints form the {\sl algebra of deformations}.
On the other hand, the TSA's first principles are (suitably general) {\sl spatio-temporal
relationalism}, which are implemented at the level of the action in ways which lead to
constraints which are then {\sl merely required to be consistent}.

Deriving GR from alternative first principles, such as those which I explore in this article
for the HKT and TSA routes, is a gradual and correctory process.
For, one is faced with
1) establishing a fully spelt-out and reasonably non-redundant set of postulates.
2) Carefully analyzing the extent to which each conventional relativity postulate arises from this set.
3) Investigating whether relativity arises alone or whether closely-allied postulates suggest
alternative theories, and whether these are plausible competitors to GR.
4) Investigating whether the new framework can accommodate a sufficient range of fundamental classical
matter to convincingly describe nature.
A further reason to study matter is that {\sl by the very essence} of the SR principles and
the equivalence principle (EP), it is meaningless to investigate whether these are recovered
within a derivation of vacuum relativity.
One must consider a sufficient variety of standard fundamental matter fields to grasp the former
and, additionally, of non-standard fundamental matter fields to grasp the latter.

For the TSA, the case-by-case analysis of 4) began with constructive methods
accommodating much standard bosonic physics (Sec 6.5).
This suffices as an arena in which to investigate (in Sec 7) the emergence of SR in the TSA
(in which SR is not presupposed, unlike in the HKT approach).
I then consider the indirect method of building TSA actions from within the
framework of presupposed split spacetime (Sec 8).
This serves to demonstrate that the TSA accommodates
a complete set of the usual fundamental fields coupled to GR (Sec 9).
I finally consider (Sec 10) whether the TSA's postulates makes stronger predictions
than conventional relativity: are any aspects of this or
of fundamental matter field theories emergent in the TSA?
This article's discussion of the whether the EP is emergent in the TSA (Sec 11)
is a substantial advance in this investigation.

\section{Postulates for the traditional route to relativity}

%

SR is built on the {\bf relativity principle}: that all inertial frames are
equivalent for the formulation of all  physical laws.
This requires the laws of nature to be invariant under a shared universal transformation group (rather
than Newtonian mechanics possessing Galileo invariance, Maxwellian electromagnetism possessing Lorentz
invariance and other branches of physics being free to possess either of these or yet other invariances).
There remains the issue of which group is shared.
Two obvious choices for this are distinguished by whether the laws
of nature contain a finite or infinite maximum propagation speed
(see e.g., Rindler, 2001).
If one postulates the existence of absolute time, the infinite choice is selected,
giving universally Galileo-invariant physics.
But if one adopts instead the {\bf (finite) maximum propagation speed postulate}:
that (without loss of generality by universality) light signals in vacuo propagate
rectilinearly with the same finite velocity $c$ at all times, in all directions, in all
inertial frames, then one obtains the universally Lorentz-invariant physics of SR.

The SR world then has no place for Newton's notions of absolute space and time because
it does not possess privileged surfaces of simultaneity.
Yet, following Minkowski (1908), the study of SR became rooted in a new geometrization:
as {\it spacetime}, a 4-d flat manifold equipped with an indefinite signature metric $\eta_{AB}$.
This indefiniteness then permits the existence of a new kind of privileged surfaces: null cones.
The physics associated with this necessarily {\sl universal} null cone structure is that
all massless particles move along the cones, while massive particles can only travel
from a spacetime point into the interior of its future null cone.
In free `inertial motion', massive particles follow timelike straight lines within the
null cone, while massless particles follow straight lines on the null cone.
Given Minkowski's geometrization,
it is then natural to implement the laws of physics in terms of 4-d spacetime tensors.

Einstein had two objections to the world picture described above.
First, his work was significantly motivated by the wish to abolish absolute structures {\sl in general},
while above one has merely traded some absolute structures for others.
Second, his desire for the relativity principle to hold
universally fails in the above SR implementation as regards gravitational physics.
While Einstein did not approach the first issue directly (see
e.g., Einstein, 1918, Einstein's foreword in Jammer, 1993 and
arguments in this respect of Barbour, 1995, 1999b), he did
indisputably find a resolution of the second issue.

Here is a contemporary account of this resolution.
Nearby freely-falling particles in a (nonuniform) gravitational field undergo relative acceleration.
Thus gravitation in general
requires that the pre-existing notion of inertial frames of infinite extent be replaced by a local notion.
The definition of this notion rests crucially on the material-independence of the
proportionality between inertial mass and gravitational mass.
This {\bf universal double meaning of mass} is one aspect of the {\bf equivalence principle (EP)}.\fn{The
next paragraph contains a second aspect.  For further aspects (not used in this paper)
and the accuracy to which all these aspects are known to hold true in nature, see (Will, 2001).}
One then adopts the {\bf identification of frames}: that the the local inertial frames of freely-falling
massive particles are to be identified with (local replacements of) the inertial frames of SR.
The universal double meaning of mass strongly suggests that gravitation could be included within
relativity by postulating the {\bf geometrical gravitation postulate} that spacetime housing
gravitational physics would not be flat Minkowski spacetime but rather a spacetime curved by the
sources of gravitation so that the following {\bf geodesic postulates} hold.
Minkowski spacetime's timelike and null lines of free motion followed by massive and massless particles
respectively are replaced by timelike and null geodesic curves.
This is due to the sources of gravitation causing the nonlocal bending of the null cone structure.
At this stage in the discussion, `geodesic' is meant in the {\it affine sense},
i.e that which is  incorporable by the mathematics of the affine connection.
The particular non-tensoriality of this object leads to the existence of coordinate systems in which
it vanishes at a particular spacetime point, which correspond to the freely-falling frame at that point.

A {\bf metric} $g_{AB}$ is furthermore postulated for spacetime.
This both incorporates a notion of length-time and is a geometrization of the gravitational field.
This metric is furthermore postulated to be {\bf semi-Riemannian}.
This attempted world description represents a simple choice of mathematics, which amounts to 1)
the preclusion of more general asymmetric or velocity-dependent\fn{These
are exemplified by the Finslerian generalization of the Riemannian metrics.}
metrics.
2) The connection associated with the metric and the aforementioned affine connection coincide, and no
further geometrical connections play any physical r\^{o}le (i.e the absence of `spacetime torsion').
This simple choice is vindicated by observations to date (Will, 2001).
As the metric is locally Minkowskian,
the {\bf local recovery of SR physics} (a second aspect of the EP) arises.
One implication of this is that GR then inherits the signature of spacetime from SR.

Next, field equations for gravitation are required.
The Einstein field equations (EFE's) (Einstein 1915, 1916) then follow from demanding
1) {\bf the GR principle}: that all frames are equivalent, embodied in spacetime general covariance
[the field equations are to be a general 4-d tensor equation].
2) {\bf That energy-momentum sources gravitation}, so this 4-tensor equation is to relate
energy-momentum ${\cal T}_{AB}$ to some geometrical (and hence, by prior assumption, gravitational)
property ${\cal C}_{AB}$ of spacetime.
As ${\cal T}_{AB}$ is conserved ($\nabla^A{\cal T}_{AB} = 0$) and (in the usual
cases) symmetric, consequently ${\cal C}_{AB}$ should also have these properties.
3) That an {\bf acceptable Newtonian limit} is recovered in situations with low velocities
(as compared to c) and weak gravitational fields (as compared to $c^2$).\fn{Not
all aspects of this limit have been established to date (Ehlers, 1987).}
A suitable candidate for ${\cal C}_{AB}$ is then well-known to be the Einstein curvature tensor
${\cal G}_{AB}$.

Additionally, the entirely mathematical {\bf Cartan--Vermeil--Weyl simplicities}
(Vermeil, 1917, Weyl, 1921, Cartan, 1922), that ${\cal C}_{AB}$ contains at most second-order
derivatives and is linear in these, were once held to be required for GR to be uniquely axiomatized.
Lovelock (1971) later showed that the assumptions of linearity and symmetry are
unnecessary in dimension $n \leq 4$ ({\bf Lovelock's improved simplicities}).\fn{Issues
of redundant structure do not end here, nor have they all been resolved yet.
For example, it is not watertight to date (Ehlers 1987, Ehlers \& Geroch, 2004)
whether the geodesic postulates are unnecessary given the EFE's.
Another issue (Ehlers, Pirani \& Schild, 1972) is whether the assumption of
semi-Riemannian geometry is unnecessary.}
Note that throughout the cosmological constant term $A g_{AB}$ is
an acceptable additional term in the EFE's.

Conventional matter can be adjoined straightforwardly.\fn{This is
subject to the caveat that re-expressing conventional matter in
curved spacetime form is ambiguous (MTW, Ehlers, 1987) e.g. due to
the possibility of coupling to spacetime curvature.
Thus, strictly, I mean that the simplest curved spacetime realizations of these matter laws,
for which there is at present no contrary observational evidence, can be adjoined straightforwardly.
When I subsequently name matter theories in this article, I implicitly
mean the simplest form of those matter theories coupled to GR.}
Some modern gravitational studies have considered adjoining non-conventional EP-violating matter
to the EFE's.
In the traditional route to GR, one can avoid such fields `by hand'
through continuing to impose the EP as a separate postulate.
In other routes, however, it is an interesting question what form the EP takes
and whether one might derive it rather than separately assume it.

\section{Many routes to relativity}

The more routes one has to a physical theory,
the more sources of insight and techniques of investigation become available,
and the more chances one has of finding a formulation that permits open problems to be addressed
(MTW, Hojman, Kucha\v{r} \& Teitelboim, 1976).
Also, different routes to an established theory may suggest different alternatives or
generalizations against which the established theory can be tested.
Wheeler advocated the position that there are many routes to GR.
Of the six routes listed in MTW, four are directly relevant to this article:
the traditional and Einstein--Hilbert action routes to the spacetime formulation,
and the routes in both directions between the spacetime formulation
and the standard geometrodynamical formulation in which spacetime is foliated by spatial surfaces.
The other two listed in MTW are also of illustrative value in the sense that
they show that GR turns up `often' from a priori distinct first principles:
the Fierz \& Pauli (1939) route via a spin-2 field in an unobservable flat background
and Sakharov's (1967) route in which gravitation is an effective elasticity of space that
arises from particle physics.

Quite a wide range of developments in GR, involving various changes of variables, splits and
discretizations, can also be thought of as routes.
Changes of variables include passing from metric to first-order
formulations such as those of Cartan (1925) (separately credited
as a route in the earlier account of Wheeler, 1968) or Palatini
(e.g., discussed elsewhere in MTW), as well as spinor, twistor and
Newman--Penrose formalisms (see e.g., Stewart, 1991, Penrose \&
Rindler, 1987).
Among these the Ashtekar variable formulations (Ashtekar, 1991) stand out for opening new
possibilities for quantum gravity.
At least some of these formulations can be cast in terms of spacetime
or in terms of a foliation by spatial hypersurfaces.
The traditional-variables version of the latter include various formulations:
geometrodynamics, the thin sandwich formulation (Baierlein, Sharp \& Wheeler 1962,
Wheeler 1964a, Belasco \& Ohanian, 1969, Bartnik \& Fodor 1993, Giulini, 1999), and several
formulations in terms of conformal geometry (York, 1971, 1972, 1999, Choquet-Bruhat,
Isenberg \& York, 2000, Pfeiffer \& York, 2003).
Also, the foliation by spatial hypersurfaces is not the only
possible sort of split dynamical interpretation: e.g., there are
also formulations built around one (Winicour, 2001) or two
(d'Inverno \& Stachel, 1978) families of null surfaces.
Discretizations may be taken as numerical approximations, but the notion of what is primary
may be reversed so that the discrete structures are fundamental and the continuum is an emergent
approximation.
These can also be treated in a spacetime form such as Regge
calculus (see e.g., MTW) or in a dynamical, spatial form such as
dynamical triangulation (see e.g., Carlip, 1998).
Additionally,  one might view as routes how the EFE's arise from the closed string spectrum
(Green, Schwartz \& Witten, 1987) and perhaps one day GR spacetime will be recoverable
from various more primitive structures (see Carlip, 2001 and Butterfield \& Isham, 2000 for reviews).

I argue that `many routes' is not a fact but a position that has to be established for each
candidate route.
For, in addition to the issues of what first principles the route
is based on and whether there are any deep-seated physical,
philosophical or mathematical structure reasons for these, there
are the issues that not all routes are rigorously established and
that some routes `do not quite lead to standard GR'.
For, firstly, not all axioms or details of GR might emerge (e.g.,
one might obtain Euclidean GR or 10-d GR, or EP violation).
Secondly, GR might not emerge {\sl alone} (this occurs e.g., in
the TSA route considered in this article).
%
In such cases, one can seek for more axioms to characterize GR, but it can be more illuminating to
consider rather whether alternative theories that arise thus can be dismissed on observational
grounds or remain plausible competitors to test GR against with future experiments.
This may one day lead to GR being supplanted, after which some of
the levels of structure the world is supposed to have may change,
e.g., if the world turns out to be string-theoretic.
Simpler examples of alternative theories in the vein of the above
include higher-curvature gravity theories, dilatonically-coupled
scalar--tensor-type theories such as Brans--Dicke theory (Brans \&
Dicke, 1961) and other EP violators (see e.g., Isenberg \& Nester,
1977a), and theories with massive or multiple gravitons.
These may be viewed as the outcome of dropping some of the mathematical simplicities in a variety of
routes to GR.
Thirdly, only a portion of GR might come out (the next section contains a good example of this).

Finally, a route to the vacuum EFE's alone does not suffice to describe the world,
as this contains matter (or at least appears to).
A credible route should also account for a sufficiently broad range of classical fundamental
matter fields to accommodate our current understanding of nature.
Such a robustness check is important in the traditional variables canonical program
(Misner \& Wheeler, 1957, Wheeler, 1962, Kucha\v{r}, 1976bc, 1977, Teitelboim, 1973ab),
in the Ashtekar variables canonical program (Ashtekar, 1991),
in the GR Cauchy problem (see Friedrich \& Rendall, 2000, and references therein) and
in the GR initial-value problem (Isenberg, \'O Murchadha \& York, 1976, Isenberg \& Nester, 1977b).
A more specific inquiry concerns {\sl how} matter is included.
Can matter can be `{\it added on}' (standard physics recovered),
adequately mimicked by vacuum GR ({\it already-unified} physics unveiled),
or does a new co-geometrization of gravity and matter arise ({\it unified} physics revealed)?
Among (already-)unified theories, one should further distinguish between total and
partial unification and between entirely classical unification and unification that
furthermore includes quantum mechanics.
Of course, what constitutes total unification depends on how
many laws of physics have so far been discovered!
E.g., many unified theories date back to when electromagnetism and
gravity were the only `understood fundamental forces'.
Examples include 1)  Weyl's theory (which, as Einstein observed, fails at the classical level through the
false prediction of light path dependent spectra, as reported in Weyl, 1918).
2) Kaluza--Klein theory (which by itself is well-known to be a quantum mechanical failure through
predicting only Planck-mass particles and not the much lighter particles observed in nature).
3) Already-unified Rainich--Misner--Wheeler theory (which is debatably a classical failure that is
worsened by newer physics, see footnote 14).
String theories include all physical interactions known today but have not as
yet contributed any specific and subsequently
experimentally-verified predictions (e.g., Smolin, 2003, Greene,
2005).

\section{Geometrodynamics}

\subsection{Motivation for, and possible limitations of, geometrodynamics}

As the dynamical viewpoint plays a widespread r\^{o}le in physics (c.f. quantum mechanics and gauge
theory), it is of considerable interest to formulate GR as a dynamics rather than as a spacetime arena.
Furthermore, this turns out to be a mathematically sensible way of interpreting the EFE's, and useful
as regards making astrophysical predictions for tight binaries composed of black holes/neutron stars
(Baumgarte \& Shapiro, 2003).

Formulating GR as dynamics is not straightforward.
For, dynamics requires a particular {\it evolution time}.
But time in GR plays both this r\^{o}le and
that of a particular choice of {\it coordinate time} on the spacetime manifold.
So making such a choice appears to clash with general covariance.
Geometrically, this element of choice is embodied in how spacetime is sliced into
a sequence of spacelike hypersurfaces of constant time.
Physically, these choices correspond to the perspectives of different families of observers.
This element of choice is not as large as might be na\"{\i}vely
supposed: predictions about a given constant $t_1$ hypersurface do
not depend on how one foliates the spacetime between a given
constant $t_0$ data hypersurface and the $t_1$ hypersurface.
This dynamics was first provided in an `extrinsic curvature' formulation by Darmois, Lichnerowicz,
Four\`{e}s-Bruhat and others [see (Stachel, 1992, Anderson, 2004c) for references].
Here I treat this dynamics rather in the {\sl canonical} or {\sl Hamiltonian} form of Dirac (1964) and
Arnowitt--Deser--Misner (1962) (ADM); see also (Wheeler, 1968, DeWitt,1967, Kucha\v{r} 1972, 1973,
1976abc).
An additional motivation for the Hamiltonian version is that it should certainly be considered in
attempts to quantize gravity, though these do run into conceptual and technical difficulties
(Kucha\v{r}, 1981, 1991, 1992, 1999, Isham, 1993).

It should be noted that in contrast to the spacetime picture of
GR, in the geometrodynamical picture it is assumed that the
emergent spacetimes are restricted to be time-orientable, globally
hyperbolic and with a globally fixed spatial topology (see e.g.,
Wald, 1984).
Are these restrictions justifiable?
Assuming that spacetime is time orientable is commonplace and `respectable' in the study of GR.
This amounts to disallowing regions of solutions which possess closed timelike curves,
which is supported by mathematical evidence of these being unstable both classically
and quantum-mechanically (the chronology protection conjecture of Hawking, 1992).
Assuming that spacetime is globally hyperbolic includes it not possessing naked singularities, and is
in line with the more general of the cosmic censorship conjectures of Penrose (1979).
Whether the universe permits topology change remains an observably unrestricted issue.

In this article I further assume that the spatial topology of the universe is compact
without boundary.
This is required in order for the constraints considered below to determine everything,
as opposed to leaving local physics vulnerable to additional influences from boundary conditions
at infinity (a situation with substantially different mathematics about which I make no claims).
These represent `closed' and `open universe' positions respectively.
In good part because of the aesthetic appeal of universe models
without such additional influences, `closed universe' positions
have been developed in a variety of contexts by e.g., Einstein
(1934, 1950), Wheeler (1964ab), MTW, Isenberg (1991) and Barbour
(1994a, 1995).
I furthermore argue that making such a choice for the spatial topology of the universe is indeed
a modelling assumption which is not observationally restricted.\fn{I
have been asked why this is not in contradiction with well-known literature (see e.g. Lachi\`{e}ze-Rey \&
Luminet, 1995) concerning restrictions on `cosmic topology' that would be implied by observing
`circles in the sky' (or other patterns).
Note that the standard mathematical meaning of spatial topology here is `topological class of a spatial
surface', which is an equivalence class under smooth deformations.
Then, an illustrative example (Fig 1b) is that surface A can be smoothly deformed into surface B
but not into surface C, so A and B have the same topology (an open one), while surface C has a
distinct topology (a closed one).
Then, by the definition of topology, all of this holds true regardless of how small the throat
aperture $d$ is.
The situation encountered in observational physics is, however, along the following lines.
Consider an observer at point $p$ in B, whose instrumentation is sensitive enough
to detect patterning up to a scale $D$.
If $d << D$, then this observed patterning is not distinguishable from that obtained by an observer
at point $q$ in C.
Thus an observer who observes this patterning cannot tell whether
(s)he lives in B or in C.
But B and C are topologically distinct; in particular one is open and the other is closed.
This illustrates that observation cannot baldly restrict the topology of the universe.
Rather, {\sl what observation restricts is a distinct property of the universe: its approximate large-scale shape}.
This issue can be viewed in two ways.
Firstly, one could say that one is studying a class of approximate cosmological models which have
sufficient side conditions to ignore small tubes, holes or twists.
For example, if one studies homogeneous and isotropic cosmologies, all points are equal so
there cannot be any such features (as these require distinguished points to {\sl be at}).
While one may be able to restrict the topology in such studies using observational data, these are
clearly studies of models which simply ignore the smaller-scale inhomogeneous features of the real
universe, thus the real universe's topology remains unrestricted.
Alternatively, a far more deep-seated modelling improvement would be to replace the standard notion
of mathematical topology by a refined scale-dependent notion which makes sense in terms of observer
probing scales.
However, such a program (an example of which is Seriu, 1996) remains in its infancy and it certainly
has not been applied yet to first-principles derivations of geometrodynamics.

My point is, rather, that given that the universe in these geometrodynamical programs is to be modelled
on a fixed topology in the standard sense, then the choice of this class to be an open one or a closed
one is a matter of taste rather than enforceable by observations.
For, the example above shows how if the universe is observed to be approximately spherical,
it can nevertheless be modelled by either an open or a closed universe, an argument which
straightforwardly extends to universes observed to be approximately any other closed shape.
Conversely, if the universe shows no signs of being approximately closed, this is also compatible
with it actually being closed but on a scale far larger than the Hubble radius of the observable
universe.}
Finally, in this compact without boundary case, topology change is forbidden by the other assumptions above
by a theorem of Geroch (1967).

\subsection{Geometrodynamics by rearrangement of the Einstein--Hilbert action of GR}

Geometrodynamics was originally obtained by rearranging the spacetime formulation.
I approach this via variational principles.\fn{Since
variational principles play an important part in the TSA discussed in this paper, I comment that
variational principles did play a r\^{o}le in the early days of GR (see Einstein 1916b, Weyl, 1921 and
references therein), while in geometrodynamics they have played an important r\^{o}le from the start
(Dirac 1964, ADM; see also Wheeler 1964a, DeWitt 1967, Kucha\v{r} 1976abc).}
Note that the EFE's also follow from variation of the Einstein--Hilbert action\fn{$g$
is the determinant of $g_{AB}$ and ${\cal R}$ is the spacetime Ricci scalar.
This is a vacuum action for ease of presentation;
adding matter poses no immediate problems (see the end of this section).}
\be
\mbox{\sffamily I}[g_{AB}]  = \int \textrm{d}^4x \sqrt{|g|}{\cal R} \mbox{ } .
\ee
Adopting this action involves presupposing semi-Riemannian spacetime.
ADM then proceeded to split this action with respect to a sequence of spatial slices.
The spacetime metric $g_{AB}$ is split according to
\be
g_{AB} =
\left(
\begin{array}{ll}
\beta_k\beta^k - \alpha^2   &    \beta_j    \\
         \beta_i            &     h_{ij}
\end{array}
\right)
\mbox{ } .
\ee
\begin{figure}[h]
\centerline{\def\epsfsize#1#2{0.4#1}\epsffile{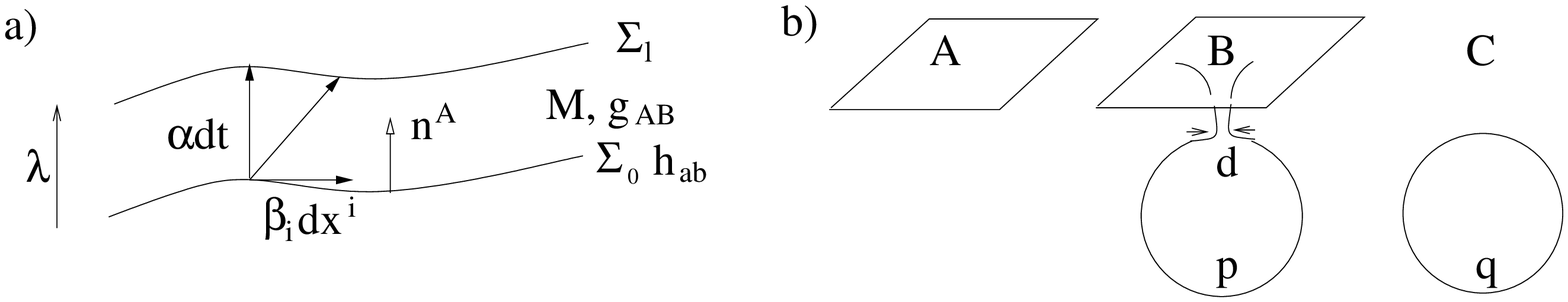}}
\caption[]{\label{TO3.ps}}
\end{figure}
Here (see Fig 1a), $h_{ij}$ is the metric induced on a spatial slice.
The {\it lapse} $\alpha$ is the change in proper time as one moves off this spatial surface.
The {\it shift}   $\beta_i$ is the displacement made in identifying the spatial
coordinates of one slice with those of an adjacent slice.    
$n_{A}$ is the normal to the spatial slice, and
\be
K_{ab} = -\nabla_a n_b  = -\frac{1}{2\alpha}\delta_{\beta}h_{ab}
\mbox{ }
\label{EC}
\ee
is the {\it extrinsic curvature} of the slice
which quantifies how bent it is with respect to the surrounding spacetime.
Here, $\delta_{\beta}$ is the hypersurface derivative, defined as the difference of
$\mbox{ } \dot{ } \mbox{ }$ (the partial derivative with respect to the label-time $\lambda$)
and $\pounds_{\beta}$ (the Lie derivative with respect to the vector ${\beta_i}$, which
represents a dragging of the surface's coordinates in the $\beta_i$ direction).

Then by standard hypersurface geometry (see e.g., Wald, 1984) the
Einstein--Hilbert action on (compact without boundary space
$\Sigma$) $\times$ (label-time interval I) is equivalent
to\fn{Here $R$ is the spatial Ricci scalar and $h$ is the
determinant of $h_{ij}$.
I drop the I and $\Sigma$ notation from the integrals from now on.}
\be
\mbox{\sffamily I\normalfont}_{\mbox{\scriptsize GR\normalsize}}[h_{ij}, K_{ij}, \alpha]
= \int_{\mbox{\scriptsize I}} \textrm{d}\lambda \int_{\Sigma} \textrm{d}^3x \sqrt{h}\alpha
(  R + K_{ij} K^{ij} - K^2  ) \mbox{ } .
\label{unellag}
\ee
One can then easily extract a split Lagrangian formulation of GR from
this by using the definition (\ref{EC}):
\be
\mbox{\sffamily I\normalfont}_{\mbox{\scriptsize GR\normalsize}}[h_{ij}, \dot{h}_{ij}, \beta_i, \alpha]
= \int \textrm{d}\lambda \int \textrm{d}^3x
\bar{\mbox{\sffamily L\normalfont}}_{\mbox{\scriptsize GR\normalsize}} =
\int \textrm{d}\lambda \int \textrm{d}^3x \sqrt{h}\alpha
\left[
R + \frac{   \mbox{\sffamily T}_{\mbox{\scriptsize GR}}   }{4\alpha^2}
\right]
\mbox{ } ,
\label{oi}
\ee
where $\bar{\mbox{\sffamily L\normalfont}}_{\mbox{\scriptsize GR}}$ is the GR Lagrangian
density and $\mbox{\sffamily T}_{\mbox{\scriptsize GR}}$ is the GR kinetic term
\be
\mbox{\sffamily T}_{\mbox{\scriptsize GR}}[h_{ab}, \dot{h}_{ab}, \beta_i] =
(h^{ac}h^{bd} - h^{ab}h^{cd})(\delta_{\beta}h_{ab})(\delta_{\beta}h_{cd})
\mbox{ } .
\label{kinterm}
\ee

The canonical form of GR is obtained from this by the procedure
standard in the principles of dynamics (see e.g., Lanczos, 1949).
Firstly, the canonical momenta are:
\be
p^{ab} = \frac{  \pa\bar{\mbox{\sffamily L\normalfont}}  }{\pa\dot{h}_{ab}} = \frac{ \sqrt{h} }{ 2\alpha }
(h^{ac}h^{bd}  - h^{ab}h^{cd})\delta_{\beta}h_{bd}
\mbox{ } ,
\label{canmom}
\ee
while $\alpha$ and $\beta_i$ are traditionally viewed as Lagrange multipliers
(coordinates whose velocities do not appear in the action), so their conjugate momenta are zero.
The action (\ref{oi}) then takes the ADM form
\be
\mbox{\sffamily I\normalfont}^{\mbox{\scriptsize ADM\normalsize}}_{\mbox{\scriptsize GR}}
= \int \textrm{d}t \int \textrm{d}^3x ( p^{ab} \dot{h}_{ab} - \alpha{\cal H} - \beta^i{\cal H}_{i})
\label{VADM} \mbox{ } ,
\ee
where
\be
{\cal H} \equiv \frac{1}{\sqrt{h}}\left(h_{ac}h_{bd} - \frac{1}{2}h_{ab}h_{cd}\right)p^{ab} p^{cd}  -  \sqrt{h}R  = 0
\label{Vham}
\ee
is the (vacuum) GR Hamiltonian constraint, and
\be
{\cal H}_i \equiv -2D_j{p_i}^j = 0 \mbox{ }
\label{Vmom}
\ee
is the (vacuum) GR momentum constraint (for $D_a$ the spatial covariant derivative).
The Hamiltonian is then
\be
\mbox{\sffamily H}^{\mbox{\scriptsize ADM\normalsize}}_{\mbox{\scriptsize GR}}
= \int \textrm{d}\lambda\int\textrm{d}^3x(\alpha{\cal H} + \beta^i{\cal H}_i) \mbox{ } .
\label{ADMHam}
\ee
As explained in Appendix A,\fn{Appendix A supplies useful
information about constraints and the Dirac (generalized
Hamiltonian) procedure.} this is the {\sl total Hamiltonian} (from
which substantial information can be extracted) as opposed to the
na\"{\i}ve Hamiltonian (which is merely zero).

Note that one has $6 \times 2$ phase space degrees of freedom per space point in the 3-metric and its
conjugate momentum, and four constraints which use up 2 of these each, so that GR has 2 configuration
degrees of freedom per space point.
The momentum constraint ${\cal H}_i$ presents no conceptual difficulties.
Via $\int - \beta^i{\cal H}_i \d^3x = \int-\beta^i(- 2D_j{p_i}^j)\d^3x =
\int - (D_i\beta_j + D_j\beta_i)\d^3x = \int -(\pounds_{\beta}h_{ij})p^{ij}\d^3x$
[where the first step uses (\ref{Vmom}), the second step uses the symmetry of $p^{ij}$ and
integration by parts and the third step uses the formula for the Lie derivative of the metric]
and the dragging significance of the Lie derivative, the presence of ${\cal H}_i$ means that true
dynamics concerns only the geometry i.e the shape of the spatial 3-surface and not the coordinate
grids painted on that shape.
Thus one can quotient (see p 10)
Riem (the configuration space of Riemannian 3-metrics on some particular fixed topological manifold
$\Sigma$) by the 3-diffeomorphisms to obtain the reduced configuration space {\it superspace}.
There remains the Hamiltonian constraint ${\cal H}$.
This contains an indefinite metric: the DeWitt (1967) metric  $G_{abcd} =
\frac{1}{\sqrt{h}}\left(h_{ac}h_{bd} - \frac{1}{2}h_{ab}h_{cd}\right)$
[the inverse of which is the $G^{abcd}$ in (\ref{canmom})], which defines an important
pointwise semi-Riemannian geometry on superspace.

\mbox{ }

\noindent{\bf {\large 4.3 Geometrodynamics from first principles?}}

\mbox{ }

Next, I consider whether the spacetime approach to relativity runs counter to
the dynamical development of

\noindent physics.
During Dirac's contribution to the above work, he did come to question the spacetime picture:
{\it ``I am inclined to believe \dots that 4-d symmetry is not a
fundamental property of the physical world"} (Dirac, 1958, p 343).
Also, Wheeler concluded his conceptual development of the Superspace viewpoint with the question:\fn{For
the purposes of interpreting his exact words, the Einstein--Hamilton--Jacobi equation is what is made
by substituting $p^{ij} = \frac{\pa \mbox{\sffamily\tiny S\normalsize\normalfont}}{\pa h_{ij}}$ into the
Hamiltonian constraint ${\cal H}$, where {\sffamily S} is Jacobi's principal function.
See e.g. (Wheeler 1968, Anderson 2004b) for a more detailed
account.
Thus the first stage of answering the question involves finding
a derivation of the specific GR form of ${\cal H}$.}
\it ``If one did not know the Einstein--Hamilton--Jacobi equation, how might one hope to
derive it straight off from plausible first principles without ever going through the
formulation of the Einstein field equations themselves?" \normalfont (Wheeler, 1968, p 273).
Could such a first principles approach to geometrodynamics be regarded as more
primary or revealing than the spacetime picture?

Note however that neither Dirac nor Wheeler subsequently made any serious attempt to free themselves of
spacetime, quite possibly due to the unresolved status of the Hamiltonian constraint.
In the case of Wheeler (1968), this manifested itself in envisaging that embeddability -- i.e. that
Riemannian 3-geometries are always to evolve such that they can be embedded in a 4-d semi-Riemannian
spacetime -- is to play a crucial r\^{o}le in geometrodynamics.
In particular Hojman, Kucha\'{r} and Teitelboim's (1976) (HKT) approach is along these lines.

On the other hand, the 3-space approach (TSA) does not presuppose this,
being based rather on relational first principles.
The nature of these means that the approach is, furthermore, a directly constructive derivation of GR
from what Barbour (1995, 1999b) considers\fn{The 1995 book cited here additionally contains others' interpretations of this.}
to be the essence of Mach's principle.
In contrast, Einstein's
traditional route does not address the abolition of absolute space directly.
It proceeds rather via

1) explaining the privileged status given to a class of locally SR
frames by the EP,

2) the mathematics of the local vanishing of the connection being
such that it is influenced by curved geometry, which is itself
influenceable by matter via the EFE's.
%
That GR is nevertheless recovered along the TSA route reconciles this issue.


As regards matter in geometrodynamical schemes, Misner and Wheeler considered the possibility
that vacuum geometrodynamics sufficed as a description of the universe, but found this to be
unsatisfactory.\fn{The hopes of this viewpoint were based on Misner \& Wheeler (1957).
There, a number of other physical features were argued to arise within vacuum geometrodynamics.
In particular, the (non-null case of the) electromagnetic field tensor was shown
to be expressible in terms of the energy-momentum tensor, from which the Einstein--Maxwell system
could be interpreted to be reducible to a (higher-order) theory of empty space alone.
This Rainich--Misner--Wheeler already-unified theory (see also Rainich, 1925 and the review Wheeler,
1962), constituted a unification of all the understood fundamental forces in its day (which just
predated widespread acceptance of the modelability of the nuclear forces by Yang--Mills theory).
The idea was then to investigate whether all other known forms of fundamental matter
could likewise be incorporated.
However, no way of incorporating spin-$\frac{1}{2}$ fermion fields was found.
%
As a result of this already in 1959 Wheeler (reprinted within Wheeler, 1962)
thought it unlikely that vacuum geometrodynamics would be sufficient to describe nature.}
One has since commonly taken geometrodynamics to be a theoretical scheme in which matter may be
`added on' (ADM, Kucha\v{r}, 1976bc).
Teitelboim (1980) succeeded in `adding on' a number of bosonic fields to the HKT scheme (see Sec 5.1),
while the TSA papers have succeeded in including a sufficient set of classical fundamental fields to
describe all of nature (see Sec 6.5 and Sec 9).
The earlier TSA papers furthermore suggested that this inclusion was perhaps more than just
`adding on' matter --  that these laws are {\sl picked out} in the TSA.
However, I have found evidence against these suggestions, by gradually finding means of {\sl also}
including nonstandard matter theories [see (Anderson, 2004b, 2005), Sec 10-11 and Appendix B].

\section{Deformation algebra first principles for geometrodynamics ?}

HKT presuppose the {\bf existence of 3-d spacelike hypersurfaces described by Riemannian geometry} and
the {\bf embeddability of these into (conventional 4-d, semi-Riemannian) spacetime}.
They implement this, as explained below, by adopting the {\it algebra of deformations} of a 3-d
spacelike hypersurface embedded in spacetime as the primary constituent of their set of first principles.
%

\begin{figure}[h]
\centerline{\def\epsfsize#1#2{0.4#1}\epsffile{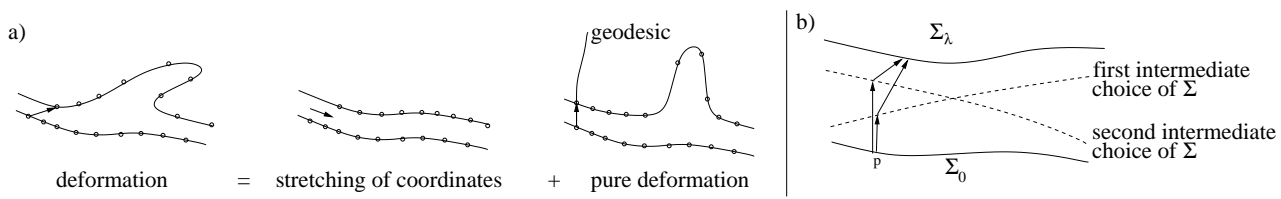}}
\caption[]{\label{1}
\footnotesize  a) Decomposition of the general deformation.  b) Composition of two deformations:
for each point $p$ on $\Sigma_0$, choose to get to a nearby $\Sigma_{\lambda}$ via either of two
intermediate surfaces.
\normalsize}
\end{figure}
%
The arbitrary {\it deformation} consists of two parts (Fig 2a): a {\it pure deformation} ${\cal P}$
along the geodesics of the

\noindent
embedding spacetime (characterized by a single proper time
function), and a {\it stretching} ${\cal S}$ of the spacelike
hypersurface's coordinates (characterized by three
coordinate functions).
A key property of spacetime as foliated by embedded hypersurfaces is the
path-independence of the foliation between two given `initial' and `final' hypersurfaces,
$\Sigma_0$ and $\Sigma_{\lambda}$.

Then (Fig 2b), by moving from $\Sigma_0$ to $\Sigma_{\lambda}$ via two distinct arbitrary
intermediate hypersurfaces (i.e subtracting what is obtained by deforming along one
of these paths from that which is obtained from deforming along the other,
i.e evaluating the Poisson brackets $\{, \}$ of deformation generators) one can obtain
(Teitelboim, 1973b) the algebra obeyed by the deformations:
\bea
\{              {\cal S}_i(x), {\cal S}_j(y)                \} =
{\cal S}_i(y)\delta_{,j}(x,y) + {\cal S}_j(x)\delta_{,i}(x,y)
\mbox{ } ,
\label{DeformAlgebra1}
\\
\{              {\cal S}_i(x), {\cal P}(y)                  \} =
{\cal P}(x)\delta_{,i}(x,y)
\mbox{ } , \mbox{ } \mbox{ }  \mbox{ } \mbox{ } \mbox{ } \mbox{ } \mbox{ } \mbox{ } \mbox{ } \mbox{ } \mbox{ }
\label{DeformAlgebra2}
\\ \{              {\cal P}(x),           {\cal P}(y)          \} =
h^{ij}(x){\cal S}_j(x)\delta_{,i}(x,y) + h^{ij}(y){\cal S}_j(y)\delta_{,i}(x,y)
\mbox{ } .
\label{DeformAlgebra3}
\eea
Here, $\delta$ is the 3-d Dirac delta function and the commas are partial derivatives.

To implement the deformation algebra as a first principle for prospective gravitational theories,
HKT evoke the {\bf representation postulate}: in order for conventional spacetime to be produced,
they demand that the ${\cal H}^{\tr}$ and ${\cal H}_i^{\tr}$ constraints for these theories
{\it close} as ${\cal P}$ and ${\cal S}_i$ do (i.e. that they form the same complete algebra).
Thus an important preliminary check is whether GR itself fits this proposal.
It does, since the deformation algebra is formally identical to the {\it Dirac algebra} obeyed by
the GR constraints.\fn{This
was originally derived by Dirac (1951) for arbitrarily-shaped spatial slices in Minkowski spacetime,
subsequently established for general spacetimes by DeWitt (1967),
and geometrically interpreted as the embeddability condition by Teitelboim (1973ab).}
The main point, however, is to {\sl allow} ${\cal H}^{\tr}$ {\sl and} ${\cal H}_i^{\tr}$ {\sl to be
a broad range of trial expressions and then see whether the representation postulate forces these
to take their GR forms} (if needs be applying subsidiary postulates).
HKT's derivation with increasingly less subsidiary postulates spanned several years and intermediate
papers (Hojman \& Kucha\v{r}, 1972, Hojman, Kucha\v{r} \& Teitelboim, 1973).
While the HKT article itself contains a subsidiary {\bf time-reversibility postulate}, the need for
this was separately removed in (Kucha\v{r}, 1974).
The remaining assumption is {\bf locality}:
the metric is to be only locally affected by a pure deformation.

The assumption of 3-d Riemannian geometries straightforwardly fixes the form of ${\cal H}_i^{\tr}$ to be the
GR ${\cal H}_i$ due to the close tie this has with the stretching of coordinates (explained on p 6).
%
This accounts for (\ref{DeformAlgebra1}) holding, as this is none other than the
composition rule for coordinate stretchings intrinsic to $\Sigma_0$.
Thus one is quickly down to Wheeler's question about the form of ${\cal H}$ alone.
That (\ref{DeformAlgebra2}) holds is also straightforward: it signifies that ${\cal H}^{\tr}$ is a
scalar density of weight 1, as can be seen by multiplying both sides by an arbitrary $\beta_i$ and
integrating over the hypersurface, giving a right hand side of the form of the Lie derivative of a
density of weight 1.

Now, while the brackets (\ref{DeformAlgebra1}, \ref{DeformAlgebra2}) that involve one or two stretches
have turned out to be both mathematically and interpretationally straightforward, the bracket
(\ref{DeformAlgebra3}) is much more involved, which is related to how the coefficients of ${\cal P}$ on
its right hand side depend on a function (the inverse, $h^{ij}$) of $h_{ij}$.\fn{As
the right hand side of a Lie algebra has constant coefficients -- {\it structure constants} -- this means that the
deformation algebra is {\sl not} a Lie algebra but a vastly more complicated structure (see Teitelboim,
1973b, p38 for a lucid account).}
A useful first step is to combine the two moves of Hamiltonian theory
$\pa H/\pa p = \dot{q} = \{q,H\}$ alongside the action of
${\cal H}^{\tr}(y)$ on $h_{ij}(x)$ being $-2K_{ij}(x)\delta(x,y)$ [in close parallel to (3)].
Thus $\delta{\cal H}^{\tr}(y)/\delta p^{ij}(x) = -2K_{ij}\delta(x, y)$, from which it follows that
any representation of ${\cal H}^{\tr}$ must be free of spatial derivatives of $p^{ij}$ ({\it ultralocal}
in $p^{ij}$), as if these were present, they would contribute derivatives of $\delta(x, y)$ but these
are absent from the equation.
HKT then set up induction proofs 1) to demonstrate that ${\cal H}^{\tr}$ must consist of a piece
quadratic in $p^{ij}$ that are in the DeWitt supermetric combination plus a (potential) piece which
is free of $p^{ij}$.
2) Additionally using {\bf Lovelock's theorem}\fn{This
is immediately tied to Lovelock's improved simplicity assumption.
Thus HKT's uniqueness is dimension-dependent -- it is false in dimension $\geq 5$
(Teitelboim \& Zanelli, 1987).}
the potential is fixed to be the $A - R$ of GR spacetime with cosmological constant $A$.
If the right hand side of (\ref{DeformAlgebra3}) is amended by a factor of -1, instead the potential
$A + R$ of Euclidean-signature GR (Riemannian rather than semi-Riemannian spacetime) emerges.
I emphasize that these proofs follow from the elements of spacetime structure present in the
mathematics of the deformation algebra.
A residual assumption that is questionable is the requirement that {\bf theories being considered
are to have precisely 2 degrees of freedom per space point}.
This is manifest in presupposing a gravitational dynamics such that

 1) the only variables are among
the 6 independent $h_{ij}$,

2) 4 degrees of freedom are then taken up by having an algebra of precisely 4
constraints ${\cal H}^{\tr}$, ${\cal H}^{\tr}_i$.
This residual assumption effectively precludes commonplace competitors such as Brans--Dicke theory
(which has an extra nonminimally-coupled scalar degree of freedom) and higher-curvature gravity
(in which higher derivatives of $h_{ij}$ are interpreted as supplementary variables).
It is only because of this assumption that HKT do not require a second-order simplicity assumption
akin to that required in the traditional route to relativity.

\subsection{Teitelboim's inclusion of matter in the deformation algebra approach}

Does the form of the Hamiltonian and momentum constraints
${\cal H}_{\mbox{\scriptsize g}\Psi}$ and $({\cal H}_{\mbox{\scriptsize g}\Psi})_i$ for GR
{\sl alongside} a range of fundamental matter fields $\Psi_{\Delta}$ fit HKT's postulates?\fn{$\Delta$
is a general multi-index which can run over a variety of spatial indices, internal indices and species
labels.}
Now, (at least in simple cases) the Einstein--matter system's Hamiltonian and momentum constraints
are of the form
\be
{\cal H}_{\mbox{\scriptsize g}\Psi} =
{\cal H}_{\mbox{\scriptsize g}} + {\cal H}_{\Psi}
\mbox{ } \mbox{ and } \mbox{ }
({\cal H}_{\mbox{\scriptsize g}\Psi})_i =
({\cal H}_{\mbox{\scriptsize g}})_i +
({\cal H}_{\Psi})_i
\mbox{ } ,
\ee
so the representation postulate idea extends additively to the constraints'
{\sl matter contributions}
${\cal H}_{\Psi}$ and $({\cal H}_{\Psi})_i$, so these {\sl separately} obey the Dirac algebra.
This property was the basis of Teitelboim's (1973ab, 1980) success in including
minimally-coupled scalars, electromagnetism and Yang--Mills theory.

This approach furthermore hinges on the extra postulate that {\bf ${\cal H}_{\mbox{\scriptsize g}\Psi}$
is ultralocal in $h_{ab}$}.
This is trivially so for the minimally-coupled scalar, while formulating an a priori unrestricted single
1-form $A_i$ ultralocally leads to the requirement that its conjugate momentum is subjected to the
emergent electromagnetic vacuum Gauss' Law.
For many 1-forms, ultralocality and the inclusion of the single 1-form subcase subject the 1-forms and
their conjugate momenta to a precursor of the Yang--Mills generalization of Gauss' Law.
This is a precursor because the internal arrays it contains have not yet been established to have the
algebraic properties which would allow these to be identified as Lie group structure constants.
Teitelboim then establishes these algebraic properties (antisymmetry in two of the internal indices
and the Jacobi identity) from the extra assumption that this emergent
constraint is to {\bf generate an internal symmetry}.
Does this extend to a cover a sufficient set of classical fundamental matter fields to accommodate
our current understanding of nature?
Calculations of mine show that there is no trouble in extending the above to cover scalar-gauge
theories, but there has been no progress with any natural inclusion of spin-$\frac{1}{2}$ fermions.
Thus the HKT route is not yet satisfactory according to the criterion of being able to include a
sufficient set of classical fundamental matter fields to accommodate our current understanding of
nature.

\section{Relationalist first principles for geometrodynamics ?}

\subsection{Relationalist first principles and their implementation}

I now move on to the second of the two first principles
approaches.
The 3-space approach (TSA) to physical systems involves
studying each system's {\it configuration space} $Q$
(i.e. the set of its canonical coordinates $q_{\Delta}$).
For example, the configuration space $Q_{\mbox{\scriptsize n}}$ for an $n$-particle system
is the space of the particles' positions.
The configuration space $Q_{\Psi}$ for a field theory with fields $\Psi_{\Delta}$
is the space of possible field values.
The configuration space Riem($\Sigma$) for geometrodynamics is the space of positive-definite
metrics on a 3-manifold $\Sigma$.

The TSA then implements two relationalist principles which are rooted in Leibniz's famous
{\sl identity of indiscernibles} (see Alexander, 1956) and which were subsequently promoted by
Bishop Berkeley (1710, 1721) and by Mach (1883).
While these ideas are sound, they suffered from a lack of theories known to implement them
until the appearance of relatively recent works (Barbour \& Bertotti, 1977, 1982).
The technique in the latter of these works is sufficiently generalizable (see also Barbour, 1994,
Pooley, 2000, RWR, Anderson, 2005) to encompass field theories in addition to particle mechanics.
The two principles are:

\noindent{\bf temporal relationalism}, that there is no meaningful
primitive notion of time for the universe as a whole.
So e.g. if all motions were to occur twice as fast, nobody would
be able to tell the difference.
Of course, one can still use one change as (perhaps a merely approximate) reference standard for
other changes.
Thus, unlike Newtonian absolute time, Leibnizian relative time remains meaningful in such schemes.

\noindent{\bf Configurational relationalism} (a generalization of spatial relationalism),
that one can declare a group of motions $G$ acting on $Q$ to be physically irrelevant.
Given $Q$, there may be a range of sensible choices for this $G$.
E.g., given $Q_{\mbox{\scriptsize n}}$, choosing $G$ to be the
trivial group gives Newtonian particle mechanics with its notion
of absolute space, while choosing $G $ to be the Euclidean group
of translations and rotations gives instead a relational
Leibnizian particle mechanics, such as in Barbour and Bertotti's
works.
The choice of $G$ as the similarity group, which includes dilatations alongside the translations and
rotations, has also been investigated (Barbour, 2003).
For $Q = \mbox{Riem}(\Sigma)$, one could again choose $G$ to be the identity, which amounts to the
coordinate grids painted on $\Sigma$ having physical significance, while the persistent relationalist
might choose $G$ to be the diffeomorphisms on $\Sigma$ to ensure that this is not the case
(Wheeler, 1968, Barbour \& Bertotti, 1982, Barbour, 1994a, RWR).
One could also choose $G$ to contain conformal transformations instead of, or alongside,
diffeomorphisms (Barbour \& \'{O} Murchadha, 1999, Anderson, Barbour, Foster \& \'{O} Murchadha,
2003, Anderson, Barbour, Foster, Kelleher \& \'{O} Murchadha, 2005, henceforth referred to as
`the conformal TSA papers').
One may view Abelian and non-Abelian gauge theory likewise: these follow from having a
configuration space of 1-forms (and perhaps associated scalars or spin-$\frac{1}{2}$ fermions),
and the choice of $G$  to be a suitable Lie group (the gauge group).
Note that throughout these examples the configurational relationalism is spatial ($G$ transforms
spatial points to other spatial points), except in the last example where it is instead
`internal' ($G$ transforms fields at each point).

Furthermore, these principles are to be mathematically implemented as follows.
Temporal relationalism is ensured by using a {\it reparametrization-invariant} action (where the
parameter is a `label time' $\lambda$).
It follows that the action is homogeneous linear in its velocities
(as then under the reparametrization $\lambda$ to
$\lambda^{\prime}$, $\int \mbox{\sffamily L}(q_{\Delta}, \frac{\d
q_{\Delta}}{\d\lambda})\d\lambda$ maps to $\int \mbox{\sffamily
L}(q_{\Delta}, \frac{\d q_{\Delta}}{\d\lambda^{\prime}}
\frac{\d\lambda^{\prime}}{\d\lambda})\d\lambda^{\prime}\frac{\d\lambda}{\d\lambda^{\prime}}
= \int \mbox{\sffamily L}(q_{\Delta}, \frac{\d
q_{\Delta}}{\d\lambda^{\prime}})\d\lambda^{\prime}$ i.e to
itself).
Then via Example 1 in Appendix A, it follows that there must be at least one relation between the
momenta i.e a {\it primary constraint}.

Configurational relationalism is to be implemented {\sl indirectly} as follows.
Write the action in the arbitrary $G$-frame by correcting the system's canonical coordinates
$q_{\Delta}$ by auxiliary $G$-variables $g_{\chi}$ (one per independent generator of transformations
in $G$).
For example, in order to render the group of translations irrelevant in particle mechanics, correct
the particle positions $q_i$ by use of the auxiliary translation variables $a_i$ to  $q_i - a_i$.
While at first sight this might look like making the theory in question into an {\sl even more}
$G$-dependent theory by passing from $Q$ to the direct product space $Q \times G$, variation with
respect to each $g_{\chi}$ then gives one {\it secondary constraint} [see Appendix A] per $\chi$,
which uses up {\sl two} degrees of freedom so that one furthermore passes from $Q \times G$ to the
{\it quotient space} $Q/G$.
As this is the space of equivalence classes of $Q$
under the motions of $G$, this approach indeed succeeds in rendering $G$ physically irrelevant.

It is worth considering some subtleties of the arbitrary $G$-frame method.
One's starting point that mathematically implements this 
is an action built from the basic objects in question 
(the $q_{\Delta}$ and various derivatives thereof) that transform appropriately between G-frames.  
These appropriate transformation properties cause 
a lot of the uses of auxiliary corrections to cancel out 
and thus not feature in the action.
E.g., for a potential built out of the objects $|q_i^{(i)} - q_i^{(j)}|$  
[where the $(i)$ and $(j)$ are particle labels], which  
is appropriate for a mechanics in which the group of translations is physically irrelevant, 
the translation-corrected versions of these objects are  
$|q_i^{(i)} - a_i  - (q_i^{(j)} - a_i)| = |q_i^{(i)} - q_i^{(j)}|$, 
which manifestly do not depend on the auxiliary translation variable $a_i$.  
Usually the only corrections which survive are those containing $\frac{\pa g_{\chi}}{\pa\lambda}$ 
which feature as corrections to the action's velocities $\frac{\pa q_{\Delta}}{\pa\lambda}$.
These arise because $\frac{\pa}{\pa\lambda}$ does not transform tensors to tensors 
under $\lambda$-dependent $G$-transformations.  
$\frac{\pa}{\pa\lambda}$ should rather be thought of as $\pounds_{\frac{\pa}{\pa\lambda}}$ 
- the Lie derivative with respect to $\frac{\pa}{\pa\lambda}$ in a particular $G$-frame - 
which transforms to the Lie derivative with respect to `$\frac{\pa}{\pa\lambda}$ corrected 
additively by the generators of $G$' [see Stewart, 1991, p. 48 for a specific example of this use 
of the Lie derivative].  
These surviving corrections are then equivalent to the `best-matching' correction terms of Barbour
and Bertotti (1982), so the arbitrary $G$-frame method constitutes a {\sl derivation} of the  
best-matching approach.
What has been gained from passing to this new perspective is

1) it has thus been proved that these corrections are homogeneous linear in velocities,\fn{It
is worth commenting that the widespread interpretation of the auxiliary variables in these
corrections is as Lagrange multiplier {\sl coordinates}.
E.g., the electric potential and the geometrodynamic shift are
usually viewed in this way.
But if these were coordinates, they would ruin the reparametrization-invariance required by the TSA.
The resolution of this issue is that they are in fact, prima facie velocities as explained in the text.
However, careful use of variational methods then reveals that the velocity and coordinate
interpretations are equivalent: both yield the same equations of motion (Barbour, 2003,
Anderson, Barbour, Foster, Kelleher \& \'{O} Murchadha, 2005).
This accounts for the acceptability of the widespread view while ensuring that the
TSA is not ruined.}
which are furthermore identified as the velocities of the actual
$g_{\chi}$ used in the method,

2) given any configuration space $Q$ and group of physically irrelevant motions $G$, 
my procedure is an explicit computation of the precise form that the best matching is to take.

The scope of this arbitrary $G$-frame method or best-matching method,\fn{I
will use these two terminologies interchangeably below rather than continually repeating that one
could consider either.} is well-illustrated by how one can obtain thus all of
Leibnizian particle mechanics models, 1-form gauge theories (Barbour \& Bertotti, 1982, Barbour,
2003, Anderson, 2004b, 2005), and spatially compact without boundary GR.
This last case is demonstrated below.\fn{This case does require adopting the convention to
use actions whose square roots are taken prior to integration, which goes beyond
the Barbour and Bertotti (1982) approach, but is discussed and resolved in (Barbour, 1994, RWR).}
Finally, I note that Pooley \& Brown (2002) and Pooley (2003-4) have argued a number of points in
favour of the above approach, for point-particle mechanics  (ontological simpleness, subsequent
extendibility to field theories, added predictivity), and also for geometrodynamics (Pooley, 2000).


This direct extraction of constraints is a key success of the above
implementation of the relational principles.
The next issue is that these directly-extracted constraints are not necessarily the only constraints 
present.  
For, as explained in Appendix A, it must be checked that known constraints propagate (the Dirac
procedure), and this can lead 1) to further constraints, and 2) onto the {\sl exhaustive
derivation of which actions within a given class of reparametrization invariant actions with a
particular group of irrelevant motions} $G$ {\sl are consistent}.
By this means, many candidate geometrodynamics were found {\sl not} to be consistent in the RWR paper,
so the TSA (essentially) singles out GR as the consistent theory.
Another point to emerge from this work is that one indeed cannot choose to associate just any
group $G$ of physically irrelevant motions with a given $Q$.
For, some choices of $G$ lead to further constraints arising from the Dirac procedure which can be
re-interpreted in terms of a larger, and enforced, group $G^{\prime}$ of irrelevant motions.
E.g., if $Q$ = Riem($M$) and one attempts to have temporal
relationalism without spatial relationalism (i.e $G$ is a priori
the trivial group), then Dirac's procedure reveals as one option
arising from the TSA for consistency that the momentum constraint
arises as an integrability condition (see also Moncrief \&
Teitelboim, 1972, 1973b, \'{O} Murchadha, 2002, and Anderson
2004a).
Then if one wishes to pursue this option (which is the one that leads to GR) one has to give up on $G$ being trivial group in favour
of it being the 3-diffeomorphisms.\fn{Thus
spatial relationalism can be considered to emerge from temporal relationalism in GR.
As this does not happen in some of the other branches of the TSA considered below,
I retain configurational relationalism as a postulate for this article.}

\subsection{That GR is a theory of the right type}

An important preliminary step is to check that GR can indeed be recast as a TSA theory.
The Baierlein--Sharp--Wheeler (1962) (BSW) formulation of GR
arises\fn{This was originally
motivated by the analogy with quantum mechanics transition amplitudes during Wheeler's attempts to
interpret geometrodynamics (Wheeler, 1964a).
It leads to the thin sandwich formulation (see p. 3 for references).}
through rewriting the Lagrangian formulation (\ref{unellag}) by solving the $\alpha$-multiplier
equation $\sqrt{h}R - \mbox{\sffamily{T}}_{\mbox{\scriptsize GR\normalsize}}/4\alpha^2 = 0$
(which is linear in $\alpha^2$) for $\alpha$ itself:
$\alpha = \frac{1}{2} \sqrt{\mbox{\sffamily{T}}_{\mbox{\scriptsize GR\normalsize}}/R}$, and using
this to {\sl algebraically} eliminate $\alpha$ from(\ref{unellag}).
Thus (assuming $R \neq 0$ everywhere in the region of interest) one obtains the BSW action
\be
\mbox{\sffamily I\normalfont}^{\mbox{\scriptsize BSW\normalsize}}_{\mbox{\scriptsize GR\normalsize}}[h_{ab}, \dot{h}_{ab}, \beta_i]
= \int \textrm{d}\lambda \int \textrm{d}^3x \sqrt{h}
\sqrt{R\mbox{\sffamily T\normalfont}_{\mbox{\scriptsize GR\normalsize}}[h_{ab}, \dot{h}_{ab}, \beta_i]}
\mbox{ } .
\label{VBashwe}
\ee
This may straightforwardly be rewritten using the TSA notation
\be
\&_{\dot{s}}h_{ab} \equiv \dot{h}_{ab} - \pounds_{\dot{s}}h_{ab}
\ee
for a metric velocity including corrections due to being in an arbitrary
$G$-frame parametrized by $s_a$ rather than the (mathematically-equivalent but
ontologically-distinct!) hypersurface in spacetime notation $\delta_{\beta}h_{ab}$ for the
metric velocity with corrections due to an arbitrary shift $\beta_a$.
Thus one has an entirely satisfactory action by the TSA criteria:
\be
\mbox{\sffamily I\normalfont}^{\mbox{\scriptsize TSA\normalsize}}_{\mbox{\scriptsize GR\normalsize}}[h_{ab}, \dot{h}_{ab}, \dot{s}_i]
= \int \textrm{d}\lambda \int \textrm{d}^3x \sqrt{h}
\sqrt{R\mbox{\sffamily T\normalfont}_{\mbox{\scriptsize GR\normalsize}}[h_{ab}, \dot{h}_{ab}, \dot{s}_i]},
\mbox{ } \mbox{\sffamily T\normalfont}_{\mbox{\scriptsize GR\normalsize}}[h_{ab}, \dot{h}_{ab}, \dot{s}_i]
= (h^{ac}h^{bd} - h^{ab}h^{cd})(\&_{\dot{s}}h_{ab})(\&_{\dot{s}}h_{cd}).
\label{TSAGR}
\ee
In this formulation, the GR Hamiltonian constraint arises as a primary constraint due to the
reparametrization invariance of this action (Subexample 2 in Appendix A),
while appropriate variation with respect to $s_a$ gives rise to the GR momentum constraint.

\subsection{Relativity without relativity}

A first distinctive feature of the RWR paper  (see also Anderson, 2003, 2004b, 2005, Kiefer, 2004)
is to {\sl furthermore not presuppose GR but rather to start with a wide class of
reparametrization-invariant actions built out of good 3-d space objects in accord with the TSA's
ontology, from which GR is derived through the overwhelming majority of the alternatives in this class
failing to lead to acceptable theories via the Dirac procedure.}

The input are ans\"{a}tze for the gravitational kinetic term
$\mbox{\sffamily T\normalfont}^{\mbox{\scriptsize trial}}_{\mbox{\scriptsize g}}$ and
the gravitational potential term
$\mbox{\sffamily V\normalfont}^{\mbox{\scriptsize trial}}_{\mbox{\scriptsize g}}$.\fn{How general
these ans\"{a}tze and their matter-field extensions really are is subject to discussion below.}
Let me take $\mbox{\sffamily T}_{\mbox{\scriptsize
g\normalsize}}^{\mbox{\scriptsize trial}}$ to be the expression
\be \mbox{\sffamily T\normalfont}^{\mbox{\scriptsize
trial\normalsize}}_{\mbox{\scriptsize g\normalsize}} =
\frac{1}{\sqrt{h}Y}G_{\mbox{\scriptsize W\normalsize}}^{abcd}
{\&}_{\dot{\mbox{\scriptsize s\normalsize}}}{h}_{ab}
{\&}_{\dot{\mbox{\scriptsize s\normalsize}}}{h}_{cd} \mbox{ } ,
\label{VASBSW} \ee which is homogeneous quadratic in the
velocities which are in arbitrary-frame form $\&_{\dot{s}}h_{ab}$,
and where \be G^{ijkl}_{\mbox{\scriptsize W\normalsize}} \equiv
\sqrt{h}(h^{ik}h^{jl} - Wh^{ij}h^{kl}) \mbox{ } , \mbox{ } \mbox{
} W \neq \frac{1}{3} \mbox{ } , \ee which is the inverse of the
most general (invertible, ultralocal) supermetric \be
G^{\mbox{\scriptsize X}}_{abcd} = \frac{1}{\sqrt{h}} \left(
h_{ac}h_{bd} - \frac{X}{2}h_{ab}h_{cd} \right) \mbox{ } , \mbox{ }
X = \frac{2W}{3W - 1} \mbox{ } . \ee I also take $\mbox{\sffamily
V\normalfont}_{\mbox{\scriptsize g\normalsize}}^{\mbox{\scriptsize
trial}}$ to be \be \mbox{\sffamily
V\normalfont}_{\mbox{\scriptsize g}}^{\mbox{\scriptsize trial}} =
A  + BR \ee which is second-order in spatial derivatives.\fn{In
fact, more general expressions for $\mbox{\sffamily
V\normalfont}_{\mbox{\tiny g}}^{\mbox{\tiny trial}}$, \ e.g.
higher curvature scalars  (the RWR paper) or ones whose
construction involves more general matrices than go into the construction of $R$
(\'{O} Murchadha, 2003), were all found to fail or to reduce to
the above form.
It should be noted however that the conventional higher-curvature theories known from the study of the
spacetime picture are precluded in the RWR paper (a point developed in Sec 10.2).}
The action is then
\be
\mbox{\sffamily I\normalfont}^{\mbox{\scriptsize trial\normalsize}}_{\mbox{\scriptsize g\normalsize}}[h_{ab}, \dot{h}_{ab}, \dot{s}_i]
= \int \textrm{d}\lambda \int \textrm{d}^3x \sqrt{h}
\sqrt{\mbox{\sffamily V\normalfont}^{\mbox{\scriptsize trial\normalsize}}_{\mbox{\scriptsize g\normalsize}}
      \mbox{\sffamily T\normalfont}^{\mbox{\scriptsize trial\normalsize}}_{\mbox{\scriptsize g\normalsize}}[h_{ab}, \dot{h}_{ab}, \dot{s}_i]}
\mbox{ } .
\ee
The above rests upon the {\bf earlier TSA papers' gravity simplicity assumption}: the pure gravity action is
constructed from a local square-root product-Lagrangian with at most second-order derivatives in
the potential, and with an ultralocal kinetic term that is homogeneously quadratic in its velocities.

Then, setting $N$ to be the emergent quantity $\frac{1}{2}\sqrt{
\mbox{\sffamily T\normalsize\normalfont}^{\mbox{\scriptsize trial\normalsize}}_{\mbox{\scriptsize g\normalsize}}
/\mbox{\sffamily V\normalsize\normalfont}^{\mbox{\scriptsize trial\normalsize}}_{\mbox{\scriptsize g\normalsize}}        }$,
the gravitational momenta are
\be
p^{ij} \equiv \frac{\partial\mbox{\sffamily{L}\normalfont} }{ \partial\dot{h}_{ij}} =
\frac{\sqrt{h}Y}{2N}G^{ijcd}_{\mbox{\scriptsize W\normalsize}}
{\&}_{\dot{\mbox{\scriptsize s\normalsize}}}{h}_{cd}
\mbox{ } ,
\label{wmom}
\ee
and (exactly paralleling Subexample 2 in Appendix A) these are related by a primary constraint
\be
{\cal H }^{\mbox{\scriptsize trial\normalsize}}_{\mbox{\scriptsize g}} \equiv
YG_{abcd}^{\mbox{\scriptsize X}}p^{ab}p^{cd} - \sqrt{h}(A + BR)  = 0
\mbox{ } .
\label{VGRHam}
\ee
Additionally, variation with respect to $s_i$ leads to a secondary constraint which is the usual
momentum constraint (\ref{Vmom}), thus ensuring that the physical content is in the shape of the
3-geometry and not in the coordinate grid painted on it.
The propagation of ${\cal H}_{\mbox{\scriptsize g\normalsize}}^{\mbox{\scriptsize trial\normalsize}}$
then gives (Anderson, 2004b, 2005)
\be
\dot{{\cal H}}^{\mbox{\scriptsize trial\normalsize}}_{\mbox{\scriptsize g}}
\approx \frac{2}{N}(X - 1)BYD_i\left(N^2 D^ip\right) \mbox{ } ,
\label{mastereq}
\ee
[where $\approx$ is Dirac's notion of equality up to weakly-vanishing terms -- see Appendix A].
The main output is the `RWR' result that the Hamiltonian constraint propagates if the coefficient
in the supermetric takes the DeWitt value $X = 1 = W$.
In this case, embeddability into spacetime is recovered.
Whether the emergent spacetime's signature is Lorentzian ($B=-1$) or Euclidean ($B =1$)
is to be put in by hand.

\subsection{GR does not arise alone in the TSA}

Note however that there are three more outputs since there are three other factors in
(\ref{mastereq}) which could conceivably vanish (RWR, Anderson, 2004ab, 2005).
1) $B = 0$, gives strong or `Carrollian' gravity options for both
$W = 1$ and for $W \neq 1$.
The former is the previously-known strong-coupled limit of GR
(Isham, 1976, Henneaux, 1979, Teitelboim, 1982, Pilati, 1982ab, 1983, Francisco \& Pilati, 1985),
a regime in which distinct points are causally disconnected by one's null cones
being squeezed into lines, which is relevant as an approximation to GR near singularities.
The latter is a similar limit of scalar--tensor theories (Anderson, 2004a).
In fact, there are two distinct versions of all of this, namely the one without a momentum constraint
which thus succeeds in being temporally but not spatially relational {\sl metrodynamics} and the one
with additionally a momentum constraint which thus are other consistent theories of geometrodynamics.
2) $Y = 0$, gives `Galilean' theories in which one's null cones
are squashed into planes and there is no gravitational kinetic term.
Strictly, for this option to arise, one should start with the Hamiltonian version of the
`Galilean' theory (its degeneracy means that there is no corresponding Lagrangian).
3) the vanishing of the fourth factor, corresponding to $p = 0$ or $p/\sqrt{h}$ = constant preferred
slicing conditions, gives a number of conformal theories
(the conformal TSA papers, Kelleher 2003, 2004, Anderson, 2005).
At least some of these can be recast by restarting with an enlarged irrelevant group $G$
that consists of the 3-diffeomorphisms together with some group of conformal transformations.
%
%
The mathematics of this option is closely related to the conformal method (Lichnerowicz, 1944,
York, 1972, 1973) for the GR initial-value problem.
Indeed, these conformal options include a {\sl derivation of GR, alongside the conformal method
of treating its initial-value problem, from a relational perspective}.\fn{It
would be interesting explore in similar spirit to this article the Wheeler--Isenberg shift of stance
(Wheeler \& Isenberg, 1979, Isenberg 1981) from geometrodynamics and the thin sandwich formulation
to the conformal geometrodynamics that follows from the conformal approach to the GR initial-value
formulation, a comparison of the various conformal GR initial-value formulation ontologies, and of
the routes leading to these from various different underlying first principles.}

\mbox{ }

\noindent{\bf{\large 6.5 Inclusion of matter in the TSA}}

\mbox{ }

The second theme of the TSA papers concerns the inclusion of fundamental matter.
This is important for the TSA, both as a robustness test and to establish
whether SR and the EP are emergent or require presupposition in this approach.

In the RWR paper, minimally-coupled scalars were successfully included alongside
demonstration that these and gravity share null cones.
There also, the inclusion of electromagnetism was demonstrated alongside the further argument
that this is emergent and uniquely picked out in the TSA.
Similarly, Barbour, Foster and \'{O} Murchadha (2002b) obtained the corresponding U(1)-scalar gauge
theory as picked out from an ansatz involving an arbitrary 1-form and two minimally-coupled scalar
fields that are allowed to interact with the 1-form, and Anderson \& Barbour (2002) obtained
Yang--Mills theory as picked out from an ansatz involving an arbitrary number of mutually
interacting 1-forms.

These constructions, all of which assume homogeneous quadratic kinetic term ans\"{a}tze,
suffice as an arena in which to investigate the local emergence of SR (Sec 7).
Using and extending the split spacetime framework techniques of Sec 8,
serves firstly to show that spin-$\frac{1}{2}$ fermions and all the 1-form
field--scalar and --fermion gauge theories can be included (Sec 9).
Thus the TSA passes the robustness test of being able to include a broad enough set of
classical fundamental matter fields to accommodate our current understanding of nature.
Secondly, it permits investigation (Sec 10) by means of a host of examples whether the suggestion that
electromagnetism and Yang--Mills theory are uniquely picked out by the TSA represents additional insight
into theoretical physics or is mostly an artifact of the ans\"{a}tze considered in the earlier TSA papers.
There I also examine speculations of the earlier TSA papers such as that the TSA {\sl ``even hints at a
partial unification of gravity and electromagnetism"} (RWR, p 3245), or that the TSA has a say in the
origin of mass (through accommodating massive scalars, but apparently not 1-forms).
In Sec 11, I provide new arguments against the assertion that in the TSA
{\sl ``self-consistency requires that any 3-vector field must satisfy ... the equivalence principle"}
(RWR, p 3217), backed up by the example in Appendix B.
%

\section{TSA: emergence of special relativity?}

In the TSA neither spacetime structure nor its locally SR element are presupposed.
The early TSA papers' inclusion of a range of standard bosonic matter fields minimally coupled to GR
provides a sufficient arena in which to investigate whether and how the local recovery of SR emerges in
the TSA.
This proceeds via the homogeneously-quadratic kinetic ansatz
$\mbox{\sffamily T}  =  \mbox{\sffamily T}_{\Psi} + \mbox{\sffamily T}_{\mbox{\scriptsize g}}$,
where the matter fields $\Psi_{\Delta}$ have kinetic term
\be
\mbox{\sffamily T}_{\Psi} =
G^{\Gamma\Delta}(h_{ij})
\&_{\dot{\mbox{\scriptsize s}}}\Psi_{\Gamma}
\&_{\dot{\mbox{\scriptsize s}}}\Psi_{\Delta}
\label{Kansatz}\mbox{ },
\ee
potential term denoted by $\mbox{\sffamily U}_{\Psi}$ and momenta denoted by $\Pi^{\Delta}$.

Then the implementation of temporal relationalism by reparametrization invariance leads to
a Hamiltonian-type constraint
\be
{\cal H}^{\mbox{\scriptsize trial\normalsize}}_{\mbox{\scriptsize g}\Psi} \equiv
\sqrt{h}(A + BR + \mbox{\sffamily U\normalfont}_{\Psi})
-
\left(
Y
G_{abcd}^{\mbox{\scriptsize X}}p^{ab}p^{cd}
+ \frac{G_{{\Gamma\Delta}}
\Pi^{{\Gamma}}\Pi^{{\Delta}}}{\sqrt{h}}
\right) = 0 \mbox{ } .
\ee
Then apply Dirac's procedure: assuming that $\mbox{\sffamily U\normalfont}_{\Psi}$ at worst depends on
connections (rather than their derivatives, which is true for the range of fields in question), the
propagation of ${\cal H}^{\mbox{\scriptsize trial\normalsize}}_{\mbox{\scriptsize g}\Psi}$ gives
$$
\dot{{\cal H}}^{\mbox{\scriptsize trial\normalsize}}_{\mbox{\scriptsize g}\Psi} \approx  \frac{2}{N}D^a
\left\{
N^2
\left(
Y
\left\{
B
\left(
D^bp_{ab} + \{X - 1\}D_ap
\right)
+
\right.
\right.
\right.
\mbox{ } \mbox{ } \mbox{ } \mbox{ } \mbox{ } \mbox{ } \mbox{ } \mbox{ } \mbox{ } \mbox{ } \mbox{ }
\mbox{ } \mbox{ } \mbox{ } \mbox{ } \mbox{ } \mbox{ } \mbox{ } \mbox{ } \mbox{ } \mbox{ } \mbox{ }
\mbox{ } \mbox{ } \mbox{ } \mbox{ } \mbox{ } \mbox{ } \mbox{ } \mbox{ } \mbox{ } \mbox{ } \mbox{ }
$$
\be
\left.
\left.
\left.
\left(
p_{ij} - \frac{X}{2}ph_{ij}
\right)
\left(
\frac{\pa \mbox{\sffamily U\normalfont}_{\Psi}}{\pa {\Gamma^c}_{ia}}h^{cj} -
\frac{1}{2}\frac{\pa \mbox{\sffamily U\normalfont}_{\Psi}}{\pa{\Gamma^c}_{ij}}h^{ac}
\right)
\right\}
+ G_{\Gamma\Delta}\Pi^{\Gamma}
\frac{\pa \mbox{\sffamily U\normalfont}_{\Psi}}{\pa(\pa_a\Psi_{\Delta})}
\right)
\right\}
\mbox{ } .
\label{sku}
\ee
This is just an extension of (\ref{mastereq}) to include some matter fields.
The terms in (\ref{sku}) are then required to vanish for consistency.
This can occur according to various options, each of which imposes restrictions on
${\cal H}^{\mbox{\scriptsize trial\normalsize}}_{\mbox{\scriptsize g}\Psi}$.
Furthermore, these options turn out to be very much connected to
the options encountered in the usual development of SR.

It should be pointed out here that the choice of a universal transformation law in setting up SR is not
merely restricted to the Galilean or Lorentzian fork that Einstein faced.
Thus the $c = 0$ Carrollian transformation option considered below is a legitimate if esoteric
posthumous prong of Einstein's fork, the possibility and dismissal of which could logically if not
historically be considered part of the mainstream derivation of SR.
This Carrollian option occurs above if one declares that $B = 0$.
The vanishing of the other factors is attained by 1) declaring that $\mbox{\sffamily U\normalfont}_{\Psi}$
cannot contain connections (see Sec 11 for a more thorough investigation of connection terms).
2) It then being `natural' for $\mbox{\sffamily U}_{\Psi}$ not to depend on $\pa_a\Psi_{\Delta}$ either
(ultralocality in $\Psi_{\Delta}$), which removes the last term.
Thus one arrives at a world governed by Carrollian Relativity.
Of course, we have good reasons to believe nature does not have $c = 0$, but what this option does lead
to is alternative dynamical theories of geometry to the usual GR geometrodynamics.
Some are spatially relational and some are not.
This is an interesting fact from a broader perspective: it issues a challenge to why the TSA
insists on geometrodynamical theories since metrodynamical theories are also possible.
But what happens in the GR option is that the momentum constraint is an integrability condition (p 11-12)
so one is stuck with geometrodynamics whether one likes it or not.

One could also enforce consistency above by the `Galilean' strategy of choosing  $Y = 0$.
This removes all but the last term.
It would then seem natural to take $\Pi^{\Delta} = 0$, whereupon the fields are not dynamical.
Moreover this does not completely trivialize the fields since they would then obey analogues of
Poisson's law, or Amp\`{e}re's, which are capable of governing a wide variety of complicated patterns.
Thus one arrives at an entirely nondynamical `Galilean' world.
In vacuo, this possibility cannot be obtained from a BSW-type Lagrangian (the kinetic factor is badly
behaved) but the Hamiltonian description of the theory is unproblematic.
Of course, the Hamiltonian constraint is now no longer quadratic in the momenta:
\be
{\cal H}^{(\mbox{\scriptsize Y = 0\normalsize})} = A + B R  + \mbox{\sffamily U\normalfont}_{\Psi} = 0
\mbox{ } .
\ee
This option is not of interest if the objective is to find {\sl dynamical} theories.
Nevertheless, this option is a logical possibility, and serves to highlight how close parallels to the
options encountered in the development of SR arise within the TSA.

There is also a combined locally Lorentzian physics and spacetime structure strategy as follows.
The signature is to be set by hand (one could just as well have any other nondegenerate signature
for the argument below).
Take (\ref{sku}) and introduce the concept of a gravity--matter momentum constraint
$({\cal H}_{\mbox{\scriptsize g}\Psi})_a$
by using $0 = -\frac{1}{2}({\cal H}_{\Psi})_a + \frac{1}{2}({\cal H}_{\Psi})_a$ and refactoring:
$$
\dot{{\cal H}}^{\mbox{\scriptsize trial\normalsize}}_{\mbox{\scriptsize g}\Psi}\mbox{$\approx$}\frac{2D^a}{N}
\left(
N^2
      \left\{
Y
             \left(
B
                   \left\{
                          \left(
\underline{
D^bp_{ab} }
-
\underline{
\frac{1}{2}
\left[
\Pi^{\Delta}
\frac{\delta\pounds_{\dot{\mbox{\scriptsize s\normalsize}}}\Psi_{{\Delta}}}
{\delta\dot{s}^a}
\right]}
                          \right)
+
\underline{
\frac{1}{2}
\left[
\Pi^{{\Delta}}
\frac{\delta\pounds_{\dot{\mbox{\scriptsize s\normalsize}}}\Psi_{{\Delta}}}{\delta\dot{s}^a}
\right]}
                   \right)
            \right\}
+ \underline{G_{{\Gamma\Delta}}\Pi^{{\Delta}}
\frac{\pa \mbox{\sffamily U\normalfont}_{\Psi}}{\pa(\pa_a\Psi_{{\Delta}})}    }
      \right.
\right.
$$
\be
\mbox{ } \mbox{ } \mbox{ } \mbox{ } \mbox{ } \mbox{ } \mbox{ } \mbox{ } \mbox{ } \mbox{ } \mbox{ } \mbox{ } \mbox{ } \mbox{ } \mbox{ } \mbox{ } \mbox{ } \mbox{ }
\left.
       \left.
            + \underline{YB(X - 1)D_ap}
            + \underline{Y
            \left(
p_{ij} - \frac{X}{2}ph_{ij}
            \right)
            \left(
            \frac{\pa \mbox{\sffamily U\normalfont}_{\Psi}}{\pa {\Gamma^c}_{ia}}h^{cj}
          - \frac{1}{2}\frac{\pa \mbox{\sffamily U\normalfont}_{\Psi}}{\pa{\Gamma^c}_{ij}}h^{ac}
            \right)    }
      \right\}
\right)
\mbox{ } ,
\ee
so that the first two underlined terms are then proportional to
$({\cal H}_{\mbox{\scriptsize g}\Psi})_a$.\fn{$\frac{\delta A}{\delta B}$
denotes the functional derivative, and the square bracket
delineates the factors over which the implied integration by parts is applicable.}
In the orthodox general covariance option, the third and fourth underlined terms cancel,
the significance of which is the enforcement of a universal null cone.
This needs to be accompanied by some means of discarding the fifth underlined term.
Here, one can furthermore {\sl choose} the orthodox option $X = 1$: the recovery of
embeddability into spacetime corresponding to GR (RWR result), or, {\sl choose} the alternative
preferred-slicing worlds of $D_ap = 0$ which are governed by equations similar to those in the
conformal initial value problem formulation of GR.
As both of these options are valid, the recovery of locally-Lorentzian physics is not a feature of
generally-covariant theories alone!
The connection terms (sixth underlined term) must also be discarded, but the Dirac procedure happens to
do this automatically for the ans\"{a}tze in question (see also Sec 11).
Thus in the TSA, locally-Lorentzian GR spacetime arises as one option, alongside
Carrollian worlds, Galilean worlds and locally-Lorentzian preferred slicing worlds.
These alternatives all lack some of the features of GR spacetime.

While the above shows that the zero, finite or infinite choice of maximum propagation speed for nature
arises straightforwardly in the TSA, and that one can choose to accompany the Lorentzian choice with
the choice for embeddability of space into GR spacetime, it is important also to mention and explore
whether the SR principle itself has a tacit influence on the above reasoning.
Could one not split the matter up and have each portion of it follow a different strategy among
the above?
From the point of  view of the conventional route to relativity this is heresy, but it is important
to consider this here to see whether the SR principle is a primary albeit hitherto tacit assumption
in the TSA or whether it {\sl emerges} from the above consistency arguments.
Now, the ultralocal and nondynamical strategies for dealing with the last term in (\ref{sku})
are available in {\sl all} the above options.
It may not be so shocking that degenerate and dual-degenerate possibilities might coexist.
Indeed if a Carrollian matter species is included in the Galilean option, then a BSW-type
Lagrangian does exist for the resulting theory.
But in the Lorentzian case this would means that the SR principle appears not to be replaced!
As things stand, one does derive that gravitation enforces a {\sl unique finite} propagation speed,
but the possibility of coexisting with fields with infinite and zero propagation speeds
is not precluded by consistency.
Thus, does the TSA not address Einstein's foundational unease over
Newtonian mechanics and Maxwellian electromagnetism having different relativities?
In other words, one might suspect that the TSA formulation is reopening the door to Aether analogues
playing a r\^{o}le in some branches of physics, unless the SR principle (which is decidedly {\sl not} a
3-space concept) is acknowledged as a hitherto tacit postulate of the TSA.

That such a dilemma exists was simply overlooked in the earlier RWR papers since it was claimed that
these ultralocal and nondynamical strategies only lead to trivial theories, in the latter case by
counting arguments.
Unfortunately, inspection of this triviality reveals it to mean `less complicated than in conventional
Lorentzian theories' rather than `devoid of mathematical solutions'.  In particular, the counting
argument is insufficient since this does not take into account the geometry of the restrictions on the
solution space.
So, while it is true that more conditions than degrees of freedom means that there is typically no
solution, some such systems will nevertheless have {\sl undersized} rather than empty solution spaces.
I provided examples of this in (Anderson, 2005).
However, it turns out that the TSA is safe from the previous paragraph's problem, as the erroneous claim
can be replaced by arguments that specific properties required for matter to be consistent by the
nondynamical strategies prevent such matter from ruining the emergence of the SR principle.
For, in the case of adjoining zero-momentum Galilean fields, these fields therefore cannot propagate,
so that it does not matter that their propagation speed is in principle infinite.
And in the case of adjoining Carrollian fields, their ultralocality precludes them from propagating
information away from any point.
There is an additional algebraic way of making the fourth underlined term vanish.
In the case of Maxwell theory, this is the parallel electric ($\underline{E} = \underline{\Pi}$)
and magnetic $\underline{B}$ field way of attaining a zero Poynting vector
$\underline{E} \mbox{ \scriptsize $\times$ }\underline{B}$ of momentum flux.
But there is then no propagation of light in this case either, as
this is maintained by mutual orthogonality in the $\underline{E}$ and $\underline{B}$ fields causing
each other to continue to oscillate, while the fields in this case are purely parallel.
In the absence of such propagation, the concept of a 1-form particle moving in a background solution
of theory T makes no sense (since this is but an approximation to the field equations of theory T,
which permit no 1-form propagation).
Thus if relativity is to be a theory about propagation speed in practice rather than in principle,
neither of the above sorts of theories affect the recovery of the SR principle from the TSA.
However such Carrollian or non-dynamical Galilean fields might still play a r\^{o}le elsewhere in
physics via coupling to conventional dynamical fields via interaction terms.
I do not know at present if this possibility can be overruled or whether it could have interesting
applications.\fn{Another approach to overcoming the abovementioned error in the earlier TSA papers,
which could also clear up this possibility, would be a rigorous proof that the incorrectly excluded
examples are too insignificant a subset of the theory's solutions to be of physical significance
(a genericity proof).}

\section{Split spacetime framework as a tool to investigate the TSA}\normalsize

This section concerns a technique which I originally employed as an indirect but convenient
way of including a full set of fields to describe nature (Sec 9).
Furthermore, I then found that details of this technique are of crucial significance in addressing
whether these fields are meaningfully picked out by the relational content of the TSA and the subsequent
issues of whether there is unification or an emergent equivalence principle (Sec 10--11).
Hence I present this technique below including these details.

The technique involves stepping outside the TSA ontology into the {\it hypersurface} or
{\it split spacetime framework (SSF)} which does presuppose the GR notion of spacetime.\fn{Thus
this is only available for coupling matter to the {\sl GR option} of the TSA.}
The point of this is that there is then a systematic treatment of Kucha\v{r} (1976abc, 1977)
by which the spacetime formulation of specific matter theories can be recast in SSF form.
I then subsequently attempt to recast this SSF form so that both a BSW-like procedure may be
followed to eliminate the lapse resulting in a {\sl reparametrization-invariant action} and that
this ensuing action is expressed {\sl in arbitrary-frame form}.
By this process, I succeed in constructing actions, which possess the two characteristics
required for them to be an acceptable starting point for the TSA, for a number of
specific matter theories known from the conventional spacetime formulation literature.
The SSF works as follows.
The presupposition of spacetime ensures that it is meaningful to decompose each matter field into
perpendicular and tangential parts with respect to an arbitrary embedded spatial hypersurface,
by use of the normal $n_{A}$ to this hypersurface and the projector $e^a_A$ onto this hypersurface.
E.g., \be A_A = n_A A_{\perp} + e^a_A A_a \label{trivia} \ee for
the 1-forms considered below.
Then $({\cal H}_{\mbox{\scriptsize g}\Psi})_{i}$ and ${\cal H}_{\mbox{\scriptsize g}\Psi}$
are decomposed into pieces which have natural interpretations from the spacetime perspective.
$({\cal H}_{\mbox{\scriptsize g}\Psi})_{i}$ consists straightforwardly of a {\sl shift kinematics}
piece which accompanies velocities and originates from Lie derivatives with respect to the
shift.\fn{This fits in straightforwardly with the r\^{o}le played by the momentum constraint
in each of Secs 4,5 and 6.}
${\cal H}_{\mbox{\scriptsize g}\Psi}$  splits into {\sl tilt} and {\sl translation} pieces,
corresponding to the further split of an arbitrary (pure) deformation of a hypersurface $\Sigma$
near a point $p$ as illustrated in Fig 3.
\begin{figure}[h]
\centerline{\def\epsfsize#1#2{0.4#1}\epsffile{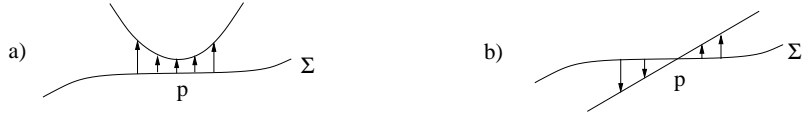}}
\caption[]{\label{TO7.ps}
\scriptsize a) The translation part is such that $\alpha(p) \neq 0$, $[\alpha_{,a}](p) = 0$.
 b) The tilt part is such that $\alpha(p) = 0$, $[\alpha_{,a}](p) \neq 0$.
\normalsize}
\end{figure}
%
\noindent The translation piece further splits into a translation on a background spacetime piece
and a {\sl derivative coupling} piece which alters the nature of the background spacetime.

Note that (Kucha\v{r}, 1976abc) the shift, derivative-coupling and tilt kinematics pieces that arise
from the presupposition of spacetime are {\sl universal}: they depend solely on the rank of the tensor
matter field rather than on any details of that particular field.
Through two of its universal features, the SSF splitting furthermore provides an interpretation of
how the TSA structures are included within the spacetime ontology.
1) {\sl Use of the arbitrary 3-diffeomorphism frame is none other than shift kinematics.}
2) {\sl The absence of tilt terms is a guarantee of an algebraic BSW procedure
leading to a reparametrization-invariant form for the action.}\fn{For,
these contain spatial derivatives of the lapse $\alpha$, which may
compromise the algebraicity of the elimination of $\alpha$ from its own multiplier equation.
It is a relevant source of worry that the tilt obstruction is a difficulty with spatial derivatives
because these are a feature of field theories but not of the particle mechanics theories that
Barbour was preoccupied with in conceptualizing and establishing the TSA (Barbour \& Bertotti,
1977, 1982, Barbour, 1994a, RWR).}
From the TSA perspective, this is an obstruction which corresponds to there not being a TSA action
for such a formulation of such a theory in the first place.
However, a number of mathematical tricks can be used toward producing tilt-free reformulations: in some
formulations of some theories tilts cancel, vanish weakly, are removable by integration by parts, or
vanish by lapse-dependent redefinition of variables.
These tricks play a crucial r\^{o}le in Secs 9--11.
The third universal feature of the SSF, derivative coupling, has no relational significance.
Note however that the absence of derivative coupling is none other than the earlier TSA papers'
gravity--matter simplicity assumption\fn{Note
that this simplicity assumption is not exclusive insofar as minimally-coupled scalar fields, Dirac
spin-$\frac{1}{2}$ fields, electromagnetism and Yang--Mills theory (and the massive and gauge theory
extensions of these last two), can all be formulated thus.}
which ensures that the matter fields do not alter the gravitational theory,\fn{In
contrast to GR in conjunction with nonderivative-coupled fields, which enjoys a configuration space
kinetic metric structure that can be decomposed into pure GR and pure matter blocks, the inclusion of
derivative-coupled fields leads to metric-matter cross-terms and matter field dependence
thoroughly infiltrating the kinetic metric.}
which is also crucial in Sec 11.

It is worth clarifying the origin and appearance of these crucial tilt and derivative coupling terms.
They emerge from the decomposition of the spacetime covariant derivatives in the action.
For example, $\nabla_bA_a$ is not merely the spatial covariant derivative $D_bA_a$ as it contains
further spacetime connection components; instead, interpreting these geometrically,
\be
\nabla_b A_a = D_bA_a - A_{\perp}K_{ab}
\mbox{ } ,
\label{Vderivproj3}
\ee
and it is such extrinsic curvature terms that constitute derivative coupling.
As another example, $\nabla_{\perp}A_{\perp}$ is not just proportional to $\dot{A}_{\perp}$ as it
likewise contains further spacetime connection components; instead, interpreting these geometrically,
\be
\alpha\nabla_{\perp}A_{\perp} = - \delta_{\beta}A_{\perp} - A^a\alpha_{,a}
\mbox{ } .
\label{Vderivproj4}
\ee
Note that this contains both the Lie derivative of shift kinematics (hidden inside the
hypersurface derivative $\delta_{\beta} = \dot{} - \pounds_{\beta}$) and a spatial derivative
of the lapse function $\alpha$, i.e a tilt.

\section{Enough fundamental matter fields can be included in the TSA}

In this first application of the SSF (Anderson, 2003), I showed that electromagnetism
admits a TSA formulation using the natural variables of the hypersurface split, whereas the massive
counterpart of electromagnetism ({\it Proca theory}) does not admit such a formulation.
This follows from this formulation of electromagnetism having a weakly vanishing tilt term
by virtue of the electromagnetic Gauss constraint law ${\cal G} \equiv D_i\pi^i = 0$
(where $\pi^i$ is the electromagnetic momentum) while the canonical mathematics of Proca theory
leaves $|\alpha_{,a}|^2$  terms lying around which are not weakly vanishing or removable by
integration by parts.
I then showed that Yang--Mills theory and the gauge theories of scalars associated with
electromagnetism and Yang--Mills theory can be cast into TSA form along similar lines.
Finally, I showed that Dirac spin-$\frac{1}{2}$ theory can be formulated so as not to contribute any
tilt, and in a style of formulation which may furthermore be combined with the formulations in the
above paragraph.
The result is a TSA formulation of gauge theories of spin-$\frac{1}{2}$ fermions associated with
electromagnetism and Yang--Mills theory (the classical theories underlying quantum electrodynamics,
quantum chromodynamics and the Weinberg--Salam electroweak unification).
These TSA formulations are now of the form
$\int\int \sqrt{h}(\sqrt{\mbox{\sffamily T}_{\mbox{\scriptsize g}}
(R + \mbox{\sffamily V}_{\mbox{\scriptsize Dirac}})}
+ \mbox{\sffamily T}_{\mbox{\scriptsize Dirac}})\textrm{d}^3x\textrm{d}\lambda$.\fn{Note
that this is not of the `square root form'
$\int\int \sqrt{h}\sqrt{\mbox{\sffamily{TV}}}\textrm{d}^3x\textrm{d}\lambda$ like in previous TSA
actions.
That the former and not the latter makes sense is clear from
$\mbox{\sffamily T}_{\mbox{\scriptsize Dirac}}$ being linear in its velocities, so that
the former but not the latter is reparametrization-invariant.
As being able to include Dirac theory is important, one should rather {\sl drop the non-relational
simplicity assumption of the earlier TSA papers that the kinetic term be homogeneous quadratic}.
This then has important consequences in Sec 10--11.}
Thus the simplest curved space realizations of a broad enough set of fundamental matter fields to
accommodate our current classical understanding of nature is successfully included in the TSA.

What should one conclude from this success?
From the SSF perspective, all that one can say is that all these matter theories {\sl happen}
to be formulable in terms of shift kinematics alone, without any need for the tilt or
derivative coupling universal features.
From the TSA perspective, however, can one argue a fortiori that nature can be described with less
structure than the SSF would na\"{\i}vely seem to suggest, i.e without having to resort to additional
{\sl non-spatial} tilt terms or {\sl complicated, counter-intuitive} derivative coupling terms?
It is then an issue of significant interest (expressed in the TSA papers) whether adopting the TSA
ontology is not only a philosophically valuable viewpoint but also a clearer and simpler route for
the study of physics.
In particular it is interesting to investigate (Secs 10--11) whether the TSA picks out the above
useful matter fields and excludes other fields which are unnecessary for explaining our observations
(be they more complicated alternative realizations or fields for which there is no evidence at all).

In favour of the SSF position, I note that Kucha\v{r} had already found the formulation of
electromagnetism mentioned above (Kucha\v{r}, 1976c), and made no big deal about it because it
did not suit the HKT program's requisite closure to reproduce the Dirac algebra.
He viewed this formulation as inconvenient, and furthermore avoidable, by adhering instead to
the somewhat more complicated formulation that is directly obtained from the canonical decomposition.
On the other hand, Teitelboim (1973b, 1980) was interested in how the Gauss constraints
${\cal G}$ for both electromagnetism and Yang--Mills theory are `pointed out' by arising
as integrability conditions.  In (the earlier TSA papers, Anderson
2003), it was also taken as a virtue that the simplified form `points out' the new constraint,
${\cal G}$, as an integrability condition.
However, I have a counter-argument to this.
Although if the constraint ${\cal G}$ is initially absent from this formal scheme it then happens to
be recovered as an integrability condition, there is no reason to expect in general a full recovery of initially
absent constraints as integrabilities (for example, in strong gravity, if the momentum constraint is
originally omitted, it is not subsequently recovered and one obtains metrodynamics rather than
geometrodynamics).
In view of this, the distinct TSA formulation of electromagnetism of the next section is more
satisfactory.

\section{Does the TSA pick out the matter fields of nature?}

In Secs 10--11 I work toward investigating the earlier TSA papers' suggestion of being more selective
than conventional physics as regards which matter fields are allowed.
The order in which issues are addressed in Secs 10--11 reflects how facts emerge as I build up
examples of TSA formulations of matter theories, while the Conclusion (Sec 12) is arranged
issue by issue rather than example by example.

I first delineate some aspects of nature which it is clear that the TSA has no explanatory power over.
As only the form of each matter field is restricted by the TSA's
consistency schemes, how many matter fields of each kind nature
possesses (e.g., why only one electromagnetic field, or why three
generations in particle physics) is not determined, nor are the
matter fields' gauge groups, nor are the magnitudes of coupling
constants in interaction terms.
The dimension of space isn't restricted either.
I also showed (Anderson, 2003) that all massless $p$-form fields arising from totally-antisymmetric
Lagrangians can be put into TSA form in direct analogy with my above formulation of electromagnetism.
Both this simple inclusion of other $p$-form fields which are not observed in nature
and the inability to pin down any of the above puzzling features about observed physics
serve as convincing pieces of evidence against the RWR
paper's speculation that the TSA hints at partial unification of gravity and electromagnetism.

I follow this up next by constructing counterexamples to the earlier TSA papers' speculations.

\subsection{A second way of including electromagnetism, its extension to Proca theory and
comments concerning formulation-dependent inclusions}

The $\alpha$-dependent change of variables from $A_{\perp}$ to  $A_0 = - \alpha A_{\perp}$ has
the effect of absorbing the tilt term since this can be factorized together with another
term to form the term $(\alpha A_{\perp})_{,a} = - A_{0,a}$.
Using this fact, I obtain a second TSA formulation for electromagnetism.
Moreover, this second approach remains valid if one includes a mass term (Anderson 2004b, 2005).
Giulini (personal communication) had already considered including Proca theory along these lines but
Barbour (personal communication) objected to this formulation on the grounds that these actions
are not entirely best matched.
But then I showed that this objection is not valid
since entirely carrying out the best matching turns out to only weakly disturb the variational equations.
That is, the fully best-matched Lagrangian's evolution equations differ from
the not fully best-matched Lagrangian's field equations only by some terms which are
functions of the constraints.
But these terms are weakly zero,
so these two Lagrangians are but reformulations of each other.
This example illustrates the following points.

This ability to include Proca theory should be viewed favourably
because it is useful.\fn{E.g., Proca theory appears in the study
of superconductivity, while my method of inclusion easily extends
to massive Yang--Mills theory which is useful to describe what the
weak bosons look like today.
It lends credibility to know that the TSA is a framework that one can continue to use in at least some
practical, phenomenological situations.}
It is then also clear that the TSA does not uniquely pick out electromagnetism among the
theories involving a 3-d 1-form.
This will be largely expanded on in Sec 10.2.
A further consequence is that the speculation in (Anderson \& Barbour, 2002) about the
origin of mass is seen to be untenable: this example and the working in
(Anderson, 2003) respectively succeed in including massive spin-1 and spin-$\frac{1}{2}$
fundamental fields into the TSA.
Also, now that I have demonstrated the inclusion of mass terms in the TSA,
I can add the values of the particle masses to the above list of things over
which the TSA has no more explanatory power than the conventional approach to physics.

That the $A_{\perp}$ formulation of Proca fails and that the $A_0$ formulation originally does
not look best matched but in fact can be made to be best matched also serves as a good
illustration that criteria for whether a matter theory can be coupled to GR in the TSA are
unfortunately rather formulation-dependent.
The TSA would then amount to attaching particular significance to finding and using
formulations meeting its criteria.
This is similar in spirit to how those formulations which close precisely as the Dirac algebra are
favoured in the SSF and HKT works.
In both cases the matter robustness requirement is, more precisely, {\sl to find at least one
mutually-compatible formulation for a satisfactory theory to represent each of the fundamental matter
fields present in our current understanding of nature}.

Considering also the criteria whereby the RMW interpretation of vacuum geometrodynamics was rejected,
it does appear that such formulation dependence is a common feature in the interpretation of
geometrodynamics.
Nor is this formulation dependence just a peculiarity of interpretations of geometrodynamics.
It also plays a r\^{o}le in the `adding on' of fundamental matter in the GR initial-value formulation
(Isenberg, \'{O} Murchadha \& York, 1976, Isenberg \& Nester 1977b) and in the Ashtekar variables approach at the classical level
(Ashtekar, 1991).
Also, there are (more mathematical) uses of spurious variables, recombination of equations such as
adding functionals of the constraints to the evolution equations and careful gauge choices, toward
casting the EFE's in special forms such as symmetric-hyperbolic for which good partial differential
equation theorems hold,
and toward finding formulations that are stable enough to support numerical relativity computations
of tight-binary black holes and neutron stars (Baumgarte \& Shapiro, 2003).

The consequence of having to deal with formulation-dependent statements is that bare statements
like `Theory X does not admit a formulation of mathematical type Y' should be doubted,
because there are very many ingenious ways of reformulating theory X toward that goal.
So, faced with statements like `mathematical type Y is claimed to act as a selector of theories of type
Z', one has little chance of rigorous proof but plenty of chances of constructing counterexamples
whereby the statement is demonstrated to be false or at least lacking in qualifying subclauses.
This should be borne in mind in the discussions below about the TSA.

\subsection{How permitting Dirac theory opens the door to many complicated theories}

The second ingredient for providing counterexamples is that it is not tenable to continue
to assume that the TSA action solely contains the square root of a kinetic term homogeneous
quadratic in the velocities.
This is both because this is a mathematical simplicity assumption which is unrelated to
the implementation of relational principles and because this assumption is insufficiently general
since it excludes the important case of Einstein--Dirac theory.
One should therefore switch to allowing completely general reparametrization-invariant actions.
These are overall homogeneous linear in their velocities.
The homogeneous linear actions built out of more general combinations of roots and sums are a useful
particular example of these.

Then the wider significance of the earlier TSA papers'
results about picking out fundamental matter fields from relational first principles is
compromised because they depend on the above additional non-relational simplicity assumption.
This is no mere logical gap because, while finding examples of formulations of further theories that
satisfy the relational principles but {\sl do not} satisfy this simplicity assumption is hard within the TSA
ontology, the SSF provides a relatively straightforward means of finding some.
Take further spacetime theories that are well-known to be consistent, and formulate them such that
there is no tilt obstruction to BSW elimination.
This sometimes produces complicated actions which are nevertheless reparametrization invariant and
built in an arbitrary-$G$ frame and hence are valid starting points for the TSA.
A good example is the electromagnetic action with an additional $A^4$ term switched on.
Perform the canonical split and pass to $A_0$ in place of $A_{\perp}$.
The $A^4$ term then contributes a $1/\alpha^3$ term so that the $\alpha$-multiplier equation is
now quadratic in $\alpha^2$ (instead of linear as in Sec 6.2).
So now its solution (via the quadratic formula) and its substitution back into the action
introduces a messy combination of roots and sums.
One can also consider switching on yet higher-order terms in $A^2$.
Likewise, it follows straightforwardly that
the next highest orders give cubics and quartics in $\alpha^2$, which produce even
more complicated combinations of roots and sums, and once one is into the quintics, an explicit formula
for the solution of the $\alpha$-multiplier equation does not generally exist.
By this stage, one has an implicit TSA form rather than just a complicated one!
(Thus, even if the BSW procedure is algebraic, it is not in general easy to solve nor
explicitly soluble in principle.)
Thus I have demonstrated that the TSA in fact admits a broad range of single 1-form theories if one pays
enough thought to what reparametrization invariant actions in general look like.
The TSA excluded these theories on simplicity grounds and not as a consequence
of implementing relationalist principles.

A second consequence of considering these more general actions is
that Barbour's (1994a) claim  that there emerges in the TSA an
`ephemeris' time expression $t^{\mbox{\scriptsize e}}$ such that 
$\frac{\pa}{\pa t^{\mbox{\tiny e}}} = \frac{1}{N}\frac{\pa}{\pa\lambda}$, 
$N \propto \sqrt{\mbox{\sffamily T}/\mbox{\sffamily V}}$ which contains all the dynamical
information about the universe is not generally true, for e.g. the
fermionic kinetic term is not included in such an $N$.

A third consequence is that while higher-curvature potentials in a homogenously quadratic action were
dismissed in the earlier TSA papers as inconsistent, one faces a difficult question of reformulability as
regards whether these theories can be accommodated within the more general classes of actions that arise
if one continues to implement the relational postulates but drops the earlier TSA papers' unrelated simplicities.

Finally, it is by now clear that the earlier TSA papers' result that gauge theories like
electromagnetism and Yang--Mills theory are picked out from among more general 1-form theories is
formulation-dependent and tied to non-relational simplicities of these papers' matter ans\"{a}tze.
Thus the emergence of these gauge theories is not a {\sl unique} picking out from relational first
principles, and hence the grounds on which a hint of unification was suggested have been dismissed.
The overall message is that {\sl the TSA, seen as what follows from implementing
relational principles without the adjunction of unrelated simplicities,
is not as restrictive as was originally hoped in the earlier TSA papers}.

\section{Does the equivalence principle emerge in the TSA?}

The RWR paper presents evidence that the equivalence principle (EP) is emergent in the TSA.
%
For, the potential ans\"{a}tze used in the RWR paper were sufficiently general to contain EP-violating
terms, yet the Dirac procedure applied to each action considered in which these were present led to an
inconsistent theory, isolating the EP-obeying theories as the only consistent theories included.
Further evidence is that all the matter fields successfully included above by indirect SSF methods obey
the EP.

Since the TSA is based on action principles, I next explain how one can tell at the level
of the action whether a theory's field equations obey or violate the EP.
Along the lines of Sec 2, coordinates can be provided at each particular point $p$ such that the
metric connection vanishes at $p$.
There is then no obstruction to curved spacetime matter field equations\fn{Note
that the gravitational field equations are given a special separate status in the EP
(`all the laws of physics bar gravity') and so do not interfere with the logic of this.}
which contain no worse than metric connection components being locally transformed into SR form.
However, no such transformation is possible if the field equations contain derivatives of
the metric connection, as is clear since the curvature tensor is built out of these
and cannot be made to vanish at a point by a change of frame.
These are theories with matter `coupled to the EFE's' in an EP-violating fashion.
I consider two ways in which derivatives of the metric connection in the field equations can arise
from actions.
The first sort is that there could already be derivatives of the
metric connection in the action e.g. in curvature--matter coupling
terms.
A standard example here is the Brans--Dicke action, which contains an $R\Phi$ term in one common
formulation for the Brans--Dicke field $\Phi$ (although Brans--Dicke theory is
`mild' in the sense that there exists a field redefinition which removes EP violation).
The second sort is that integration by parts could be required during the variational working causing
mere metric connections in the action to end up as derivatives of metric connections in the field
equations.
This is exemplified by most of the cases of the action
\be
\mbox{\sffamily I}[g_{AB}, A_A] =
\int \textrm{d}^4x\sqrt{|g|}
\left[
{\cal R} + \mu
\left(
               C_1\nabla_A A_B\nabla^AA^B
  +            C_2\nabla_A A_B\nabla^BA^A
  +            C_3\nabla_A A^A\nabla_BA^B
\right)
\right]
\mbox{ }
\label{NVA}
\ee
($\mu$ is here a coupling constant).
Note however that the `Maxwellian curl combination' ($C_1= -C_2$, $C_3 = 0$ case) contains no
metric connections through cancellation by antisymmetry.
(This sort of cancellation also occurs in Yang--Mills theory and in the various scalar and fermion
gauge theories associated with electromagnetism and Yang--Mills theory).
Another illustrative example is Dirac theory.
The action here does contain metric connections as part of covariant derivatives,
but these derivatives occur {\sl linearly} so there is never any integration
by parts that turns these into metric connection derivatives in the field equations.
Thus all these well-known matter fields obey the EP.

Does the TSA exclude all EP violation of the first sort?
Consider Brans--Dicke theory.
The RWR paper includes Brans--Dicke theory, but only under the field redefinition by which
its EP violation is removed.
However, I showed that the original, EP violating form of Brans--Dicke theory
can also be included in the TSA (Anderson, 2004b).
Could the TSA be salvaged from this by insisting that the earlier TSA papers' tacit,
non-relational `gravity--matter simplicity assumption' be part of the axiomatization?
(For, this disallows metric--matter kinetic cross terms and matter field dependence of the kinetic
matrix, which are features possessed by the original formulation of Brans--Dicke theory.)
However, while the `gravity-matter simplicity assumption' is an additional extraneous mathematical assumption
in the TSA, from the SSF perspective it is highly significant as the turning off of the third
universal feature: derivative coupling.
And, in the HKT approach, the Hamiltonian counterpart of this statement appears as an extra
matter postulate which is clearly identified as the {\bf geometrodynamical EP}!
Thus the RWR paper's claim of emergence of the EP is marred because the kinetic term ansatz chosen
there is built respecting a tacit simplicity assumption which complies with the EP.
While, without this simplicity assumption, the TSA admits the original formulation of Brans--Dicke
theory which is an EP violating theory of the first sort.

It is then interesting to investigate whether the TSA excludes all EP violating theories of the
second sort.
The SSF provides a useful guide in this study.
Consider the earlier TSA paper's EP violating potential terms.
In the SSF, these require partnering with certain other terms in order to form a consistent theory.
The suspicion is that these include derivative-coupled kinetic terms which are missing in the original
RWR paper through the overly restrictive form of the kinetic term ansatz, a fault that one would seek to
remedy by simply adding terms of the sorts known to be required in the SSF to form an enlarged kinetic term ansatz.
Recall however that if these certain other terms include tilt terms, then there is an obstruction to
writing down a TSA action in the first place, so that such formulations of such theories are excluded
regardless of whether there is a fault in the generality of the earlier TSA papers' kinetic term ansatz.
Indeed this turned out to complicate my investigation (Anderson, 2003)
of theories along the lines of (\ref{NVA}) in the $A_{\perp}$ formulation.
For, the actions for these {\sl do} have unpleasant extra pieces not included in the TSA 1-form
kinetic ansatz (\ref{Kansatz}) such as metric-matter velocity cross-terms and matter field dependence
in the kinetic matrix, and linear terms which cause the action to be a more
complicated combination of roots and sums.
But, throughout these examples these new terms {\sl are} partnered by tilt terms,
so these formulations of these theories cannot be cast into TSA form.

In (Anderson, 2005), I commented that this SSF working was suggestive that the unexplained magic
whereby the RWR paper excluded EP violating theories could be elevated to a `half-argument' as follows.
Firstly (Sec 8) tilt and derivative coupling terms originate from spacetime metric connection terms
the original presence of which would {\sl usually} indicate EP-violation (exclude the trivial case of
no genuine dependence by cancellation and the linear case).
This will tend to justify the metric--matter simplicity assumption in the earlier TSA papers after all,
due to extra derivative-coupling terms and extra tilt terms arising alongside each other from
connection terms, and the tilt terms then obstruct the BSW passage, thus preventing TSA
actions from being constructible in this way.

Yet, there is no good reason for nontrivial tilt and derivative coupling to
{\sl always} arise together in formulations of 1-form theories.
So I acknowledged in (Anderson, 2005) that there is in principle scope for TSA-formulable
EP violating theories of the second sort, although I had found no examples of such.

In the current article however, I break this deadlock by constructing a nontrivial example of such by
exploiting tilt-removing rearrangements on a carefully-selected massive counterpart of one of the
theories following from (\ref{NVA}), namely $(\nabla_AA^A)^2 + m^2A^2$ theory coupled to GR.
I provide details of this counterexample and of unpleasant features it possesses in Appendix B.
{\sl The traditional and HKT routes to GR avoid being coupled to these theories with unpleasant
features by having a separate EP postulate.}
{\sl The TSA does no better: if considered without a separate EP postulate,
by the counterexamples of Brans--Dicke theory and $(\nabla_AA^A)^2 + mA^2$-theory,
EP-violating theories of neither of the two sorts considered can be excluded.}
As one of the latter example's unpleasant features also compromises the emergence of locally SR
physics, I suggest the TSA position should be to acknowledge the need for supplementation by a
clearly-identified, non-relational EP postulate (rather than trying to cope with such unpleasant
theories in the absence of such a postulate).

\mbox{ }

\noindent{\bf{\Large 12 Conclusion}}

\mbox{ }

\noindent GR may be reached along many routes.
Searching for such routes can give rise to new techniques,
and to new insights into the meaning of GR, which may suggest a range of different generalizations
or alternatives

\noindent against which GR may be tested both conceptually and experimentally.
Two respects in which the formulation of GR
which emerges from Einstein's traditional route to GR
has been contended to be unsatisfactory are:

A) that a dynamical rather than spacetime theory or formulation would
fit in better with the development of the rest of physics.

B) It is not manifestly a relational formulation since it does not directly approach
what Barbour considers to be the essence of Mach's principle:
the abolition of absolute space and time required by Leibniz' identity of indiscernibles.

However, the Einstein field equations (EFE's) of GR may be rearranged as a dynamics of 3-geometries
(geometrodynamics).
This partly resolves A): spacetime can be {\sl reformulated} in terms of dynamics.
This is futhermore useful in the mathematical study of the EFE's, in large gravitational field
astrophysical applications and in the development of some approaches to quantum gravity.
One then encounters Wheeler's question: can geometrodynamics furthermore be regarded as primary,
resting on plausible first principles of its own rather than on rearrangement of the EFE's?
The Hojman--Kucha\v{r}--Teitelboim (HKT) approach and 3-space approach (TSA) provide two
ontologically-distinct answers to this.
HKT's first principles are embeddability of the spatial 3-geometry into conventional spacetime,
implemented by demanding that the constraints of a prospective gravitational theory close as
the algebra of deformations.
The TSA does not presuppose embeddability into spacetime.
Rather, its first principles are temporal and configurational relationalism,
which are implemented by considering reparametrization invariant actions for which
a group of transformations acting on the configurations is not physically relevant.
Both of these implementations lead to constraints.
The extra structure of GR then {\sl emerges} by applying the Dirac procedure to these constraints.

By the form of its first principles and the directly constructive nature of its route,
the TSA embodies what Barbour considers to be the essence of Mach's principle.
That GR arises in this way resolves B).
{\sl GR is Machian in Barbour's} (1995, 1999b, the RWR paper)
[{\sl or Wheeler's}] (1964ab, MTW) {\sl sense in addition to being a spacetime theory.}
This is clearly of philosophical interest.  Note alongside this the following caveats.

1) `GR' here means the spatially compact without boundary portion of the globally hyperbolic GR spacetimes.
While the adoption of such a portion is also motivated by Machian considerations
(see Wheeler, 1964ab), it is not free of controversy.

2) The emergence of GR in the TSA from applying the Dirac
procedure to a range of TSA ans\"{a}tze is {\sl only one of
several options}, while in the HKT approach one gets back just GR.
This is a symptom of the TSA assuming less structure than the HKT approach
though adopting space rather than spacetime.
This has the positive aspect that the other TSA options offer a glimpse of
several geometrodynamics without spacetime.
It remains to be seen whether the TSA will additionally deliver widely-applicable theoretical
insights or alternative theories that are observationally viable and experimentally testable.
At this stage it is worth noting that applications considered (Barbour 1999a, the conformal TSA papers,
Anderson 2005)
include obtaining a better understanding of the GR initial-value formulation, obtaining alternative
conformal theories of gravity, and the tentative hopes concerning quantum gravity at the end of this
Conclusion.

3) The TSA and HKT routes to GR do no better than the traditional route as regards
using mathematical simplicities unrelated to those theories' `true principles' so as to
exclude a number of other theories of gravity.
Both the TSA and the traditional route
place limitations on orders of derivatives to be considered, while the HKT approach
places the restriction that the theory of gravity is to have two degrees of freedom,
which both overrules higher derivative theories and Brans--Dicke theory.

4) As regards the level of structure in question, the HKT versus TSA debate about
geometrodynamics is about ``spacetime or space" rather than ``absolute or relative motion".
The latter involves considering whether {\it substantivalism}
(that a containing structure for physical objects should have a
similar status to the objects themselves) is vindicated, but in
both the spacetime and space viewpoints, a certain level of
reality {\sl is} ascribed to a containing structure (see e.g.,
Stachel, 1989, Jammer, 1993), whose physical r\^{o}le is the
inclusion of the modern concept of gravitation.
In this sense, both HKT and the TSA do have a substantival element.
But in neither case is it substantival in the sense that reality is ascribed to
a collection of points on the spacetime or space manifold.
Rather, reality is ascribed to the {\sl geometry} associated with the spacetime or space manifold.
Such viewpoints, which in a sense represent a compromise between
absolute and relative motion, constitute `sophisticated
substantivalism' (see e.g., Pooley, 2000), and the spacetime or
space debate is thus about two brands of this.

\mbox{ }

Many issues concerning the interpretation of geometrodynamics are meaningless without
the inclusion of matter.
Firstly, matter is required to discuss the emergence of the SR postulates in the TSA (in the HKT
approach, local SR is just a simple consequence of presupposing spacetime).
%
%
I have found that the Lorentzian--Galilean options which emerge along the traditional
route to SR also emerge as options under which the TSA mathematics is consistent.

Secondly, the inclusion of matter is a robustness test for the HKT and TSA routes.
The TSA passes the robustness test: a sufficient set of classical fundamental matter fields
can be included so as to accommodate our current understanding of nature.
The HKT approach is not known to pass the robustness test: a means of accommodating
spin-${\frac{1}{2}}$ fermions has not been found.
As regards simple adding on of matter, the ultralocality in the metric and internal symmetry generation
postulates that Teitelboim used to include the bosonic fields look less well-suited to a subsequent
fermionic extension.
While Teitelboim (see e.g,. Teitelboim, 1980) did manage to
accommodate spin-3/2 fermions (and consequently supergravity
theory) in a way that involves more structure than merely adding
on (akin to Dirac's derivation of the Dirac equation from the
Klein--Gordon equation), he comments on the lack of a known such
move to underly Einstein--Dirac geometrodynamics.
Indeed, were such a move found, that would even resolve the most major
objection to the Rainich--Misner--Wheeler interpretation!
On the other hand, this paper's emphasis on the large scope of reformulational flexibility
suggests that it would be very hard to dismiss the possibility of ever being able to find such
a formulation of Einstein--Dirac theory.

Thirdly, there are the further questions of whether the TSA is in any way selective of matter fields.
For, despite assuming less structure, the earlier TSA papers suggested that the TSA was
restrictively picking out which matter fields it can accommodate.
However, I have argued against the results that electromagnetism, Yang--Mills theory and U(1)-scalar
gauge theory are {\sl uniquely} picked out, in the sense that I have shown that the restrictiveness
of these results originates not just from implementing the relational postulates that the TSA is
supposed to have but also from imposing additional unrelated simplicities.
I have shown this by providing examples of formulations for further matter theories for which the
relational principles still hold, including some for which
the earlier TSA papers' simplicities do not hold.
The simplicities in question are the exclusion of derivative-coupled kinetic terms and
the supposition that the action is not only homogeneous linear in the velocities but
furthermore attains this by being the square root of the sum of squares of velocities.
Additionally TSA-castability of a theory turns out to be formulation-dependent.
Taking these things into account, and taking the TSA to involve the implementation of relationalist
first principles {\sl alone}, I find, contrary to the earlier TSA papers' speculations, that gauge
theory is not uniquely picked out, there is no hint of unification, account of the origin of mass,
or emergence of the equivalence principle (EP).
So in fact the assumption of less structure in the TSA yields what one would expect: more rather than
fewer possibilities are available (more theories of gravity, without there being any mysterious
restrictions lessening what kind of matter fields can be coupled to those theories).
The assumption of less structure also makes proving things harder in the TSA than in HKT.
E.g., HKT's induction proofs mentioned in Sec 5 rely on the
assumption of spacetime structure.
What {\sl is} available for the TSA is disproof by counterexample, as used in this article.

It is worth commenting further about how gauge theory emerges in the earlier TSA papers' and
Teitelboim's (1973ab) interpretations of geometrodynamics with matter.
This is a substantially different route from the conventional emergence of gauge theory,
where local symmetries are imposed on a flat spacetime background.
I suggest that this alternative route works as follows.
If one begins by accepting that one lives within a curved GR spacetime (which includes the EP holding
perfectly, or at least to very great experimental accuracy), then although non-Maxwellian combinations of
derivative terms in flat spacetime are just as Lorentz-invariant as the Maxwellian curl combination,
these other combinations {\sl could not arise as local laws in the first place since they are EP violators}.
Note that one then gets both unbroken and broken gauge theory as logical possibilities,
including unusual additional breakings by $A^4$ and $A_AA_B\nabla^A A^B$ terms in the action.
Moreover, geometrodynamical formulations are not special in this respect; rather, there should be a
{\sl formulation-independent} tie between the EP and such an emergence of gauge theory.
This will be explored elsewhere.

\mbox{ }

How successful are each of the formulations of geometrodynamics considered in this paper?
The TSA has the advantage that it is a simple procedure, and remarkably successful at accounting for our
current understanding of nature via the simplest curved-spacetime realization of the classical
fundamental matter fields without needing to know about any other supposedly universal features of
spacetime.
On the other hand, it does not tell us more than the usual approach to physics
about why these fundamental matter laws appear to be present in nature.
And detailed investigation of this particular issue benefited a great deal from
my application of another view of geometrodynamics, the split spacetime framework (SSF).

The SSF has three universal features for tensor fields:  shift kinematics, tilt and derivative coupling.
Each of these are generally requisite for consistency and, advantageously, they are
systematically computable following Kucha\v{r}.
The SSF is useful in my study, firstly, because it affords an interpretation of how the TSA works.
For, shift kinematics is none other than the use of the arbitrary 3-diffeomorphism frame, while it is the
absence of tilt that permits an algebraic Baierlein--Sharp--Wheeler (BSW) elimination procedure that
leads to a reparametrization-invariant action which could be reinterpreted as a candidate TSA action.
Additionally, the earlier TSA papers' matter kinetic term ans\"{a}tze are devoid of derivative-coupled
terms so that the third sort of universal terms requisite in the SSF are absent via an additional,
non-relational simplicity assumption.

Furthermore, this simplicity assumption is closely linked to HKT's presupposition of a geometrodynamical
version of the EP.
Thus, the RWR paper's assertion of EP {\sl emergence} is highly questionable.
Moreover, if one does rather adopt an additional geometrodynamical EP postulate,
it looks extraneous from a TSA perspective, while from a SSF perspective the geometrodynamical EP,
shift and tilt all concern the small set of universal features of the approach.

What the SSF does not offer however is an explanation of why the usual matter fields can be formulated
free of tilt, while this is a simple consequence of temporal relationalism in the TSA.
Also it is in favour of the TSA that it succeeds in accommodating the standard matter fields from
fewer assumptions of structure than the SSF requires.
That the SSF perspective readily permits the `turning on' of additional universal structure is in one
sense useful since it readily provides the means of finding and studying some alternative theories not
readily picked up in the TSA.
However, in the TSA one could choose options other than GR as a means of finding and studying other
alternative theories which are not picked up at all in the SSF.
{\sl The TSA provides new theories that are Machian dynamics of 3-geometries but which do not have a
conventional 4-d GR-like spacetime interpretation.}
Thus the spacetime and space interpretations of geometrodynamics are both interesting in their own
different ways.

\mbox{ }

I finish with comments about quantization.
The spacetime and space points of view may suggest distinct approaches to
quantum gravity and its problem of time.
Internal time (Kucha\v{r} 1992, Isham, 1993), various histories approaches (Hartle, 1995,
Savvidou, 2005), spin foams (Perez, 2003)
and causal sets (Sorkin, 2003) are denizens of spacetime thought.
Space-based approaches include the na\"{\i}ve Schr\"{o}dinger
interpretation (Hawking, 1984, Hawking \& Page, 1986, 1988, Unruh
\& Wald, 1989), Barbour's perspective (Barbour 1994b, 1999a), the
conditional probabilities interpretation (Page \& Wootters, 1983)
and the semiclassical scheme e.g. reviewed in Kiefer (2004) (also
see Kucha\v{r}, 1992, 1999, Isham, 1993, Butterfield, 2002,
Smolin, 2000 for comments and criticisms about these various
approaches).
While the spacetime perspective is a very fruitful approach to classical GR and in quantum particle
physics in a Minkowskian background, could these successes be bedazzling us and consequently misleading
us in our attempts toward quantum gravity?

Is the classical TSA of use in the search for quantum gravity?
While Barbour has a number of nice ideas as regards quantum gravity (Barbour 1994b, 1999a, Anderson 2004b),
there are difficulties in practice with connecting the classical TSA ideas with these.
For example, while in homogeneous quadratic mechanics one is to exploit the nice Riemannian geometry of
the configuration space, in contrast the BSW action does not give a nice geometry at all (Anderson, 2003)
(this is the full geometry of superspace whereas people have hitherto used instead the pointwise
geometry of superspace).
However, the conformal way of recovering GR in the TSA may have a
small chance of offering progress toward quantum gravity,\fn{This
approach moreover closely resembles an emergent York internal time
approach (see e.g., Isham, 1993) despite its timeless origin,
which in my opinion is an example of why perhaps the distinction
between spacetime and space approaches to quantum theory is not so
strong.} as might some alternative conformal theories (were these
to be established as viable classical theories first).
One sense in which reconnection may be possible is via the TSA helping consolidate how certain
closed-model interpretations of quantum mechanics suggested for use in quantum cosmology can be
seen to be made compatible with relativity.
That is work in progress (including the toy model approach in Anderson, preprint, 2006a, 2006b).

\mbox{ }

\noindent{\bf Acknowledgments}

\mbox{ }

\noindent I thank the following.
Dr. Julian Barbour for providing me with an interesting line of study and for collaborations.
Him and Prof. Don Page for substantial discussions and comments on manuscripts.
My other collaborators on these issues: Mr. Brendan Foster, Dr. Bryan Kelleher and Prof. Niall \'{O} Murchadha.
The Barbour family for hospitality during some of these discussions and collaborations.
The 3 anonymous referees for comments leading to substantial improvements in this article.
Dr. Harvey Brown for discussions and encouragement, particularly toward tackling the equivalence principle issue.
Prof. James Nester for discussions and encouragement, particularly concerning various unusual field theories.
Dr. Oliver Pooley for discussions, and both him and Dr. Jeremy Butterfield for inviting me to speak at
the Oxford Spacetime conference where I enjoyed discussions with many of my fellow participants.
Professors Malcolm MacCallum, Christopher Isham, James Vickers, Reza Tavakol and Jurgen Ehlers,
for comments, discussions and encouragement.

I acknowledge funding support over the years covered by this work at various stages by
PPARC, Peterhouse Cambridge and the Killam Foundation.

\mbox{ }

\noindent{\Large{\bf Appendix A: Constraints and the Dirac procedure}}

\mbox{ }

\noindent Let $q_{\Delta}$ be canonical coordinates labelled by some set $\Delta$.
Consider the Lagrangian $\mbox{\sffamily L} \normalfont (q_{\Delta},
\dot{q}_{\Delta})$ for some candidate

\noindent theory of the $q_{\Delta}$.
If $\mbox{\sffamily L} \normalfont$ is such that not all the conjugate momenta
$p^{\Delta} \equiv \frac{\partial \mbox{\sffamily \scriptsize L \normalsize\normalfont}}
                   {\partial\dot{q}_{\Delta}}$
can be inverted to give the velocities $\dot{q}_{\Delta}$ as functions of the $p^{\Delta}$,
then the theory has some set $\Pi$ of {\it primary constraints}
${\cal C }_{\Pi}(q_{\Delta}, p^{\Delta}) = 0$ solely by virtue of the form of {\sffamily L}.

\noindent Example 1 (Dirac): suppose the Lagrangian is homogeneous linear in its velocities.
Then momenta obtained by differentiation of the action with respect to the velocities are
homogeneous of degree 0 in the velocities.
Thus they are functions of ratios of velocities.
But only $n$ -- 1 of these ratios are independent.
Thus there must be at least 1 relation between the momenta, i.e a primary constraint.

\noindent Subexample 2 (Barbour):  if such a Lagrangian is the square root of the sum of squares of
velocities, the explicit working that determines the form of the only primary constraint follows the
form as exemplified below by the working by which the Hamiltonian constraint arises
from the BSW action of GR (\ref{VBashwe})
\be
\begin{array}{l}
G_{\mbox{\scriptsize \normalsize}ijkl}p^{ij}p^{kl}  =
G_{\mbox{\scriptsize \normalsize}ijkl}
G_{\mbox{\scriptsize \normalsize}}^{ijcd}
\sqrt{           \frac{  R   }{   \mbox{\sffamily{\scriptsize T}}_{\mbox{\tiny GR}}      }          }
(\delta_{\beta}h_{cd})
G_{\mbox{\scriptsize \normalsize}}^{klab}
\sqrt{           \frac{   R  }{   \mbox{\sffamily{\scriptsize T}}_{\mbox{\tiny GR}}      }          }
(\delta_{\beta}h_{ab})
= \sqrt{h}\frac{    R    }{    \mbox{\sffamily{\scriptsize T}}_{\mbox{\tiny GR}}    }
\mbox{\sffamily T}_{\mbox{\scriptsize GR}} = \sqrt{h}R
\end{array}
\ee
by the definition of momentum, and using that $G_{ijkl}$ is the inverse of $G^{ijcd}$.

\noindent Furthermore, there may be {\sl secondary constraints} ${\cal C}_{\Sigma} = 0$ among the
variational equations (the non-constraint variational equations are {\sl evolution equations}).
Let ${\cal C}_{\Gamma}$ denote the ${\cal C}_{\Pi}$ and ${\cal C}_{\Sigma}$.

All one has done so far is to show that these constraints hold
for some initial (label)-time configuration.
But it should not be the case that some configuration has the privileged status that constraints hold
for it but not for other configurations corresponding to other values of the (label-)time.
What one requires is Dirac's (1964) generalized Hamiltonian procedure.
One way to phrase this (used in the TSA) is to check whether the ${\cal C}_{\Gamma}$ are propagated
away from the `initial (label)-time' configuration by the theory's evolution equations.\fn{Another
way of phrasing this is in terms of the Poisson bracket algebra of the known constraints
(as used by HKT).}
If $\dot{{\cal C}}_{\Gamma}$ is some functional of the ${\cal C}_{\Gamma}$,
then they are propagated, since we already know that each of the ${\cal C}_{\Gamma}$ is 0.
This vanishing up to (already-known) constraints is termed {\it vanishing weakly}
and is denoted by $\approx 0$.
Then one says that the existing constraint algebra of the ${\cal C}_{\Gamma}$ {\sl closes}.
For example, the ${\cal H}$ and ${\cal H}_i$ of GR close at this stage
to form the Dirac (or deformation) algebra.
However, it can happen that $\dot{\cal C}_{\Gamma}$ does not vanish weakly,
i.e. it contains further independent functions expressions ${\cal C}_{\Sigma_{2}}$,
which are then themselves required to vanish in order for the theory to be consistent.
In this case one has found more constraints.
One must then enlarge $\Gamma ( = \Gamma_1)$ to $\Gamma_2$ which comprises
both $\Gamma_1$ and $\Sigma_2$.
In principle, this becomes an iterative process by which one may construct a full
constraint algebra ${\cal C}_{\Gamma_{\mbox {\scriptsize final \normalsize}}}$
by successive enlargements to ${\cal C}_{\Gamma_{i+1}}$
comprising ${\cal C}_{\Gamma_{i}}$ and ${\cal C}_{\Sigma_{i+1}}$.

In practice, however, the process cannot continue for many steps since
for the process to continue, at each step the number of independent ${\cal C}_{\Gamma_{i + 1}}$
must be greater than the number of independent ${\cal C}_{\Gamma_{i}}$.
But each constraint uses up further degrees of freedom, so if there is only some finite number
$\Delta$ of canonical coordinates available, a long enough Dirac procedure will render the
prospective theory inconsistent, trivial
or with an insufficiently sized solution space to be a realistic theory.
That some candidate Lagrangians thus lead to inconsistent theories should not come as a shock --  there
is no guarantee that a given Lagrangian will give rise to any consistent theory, as is illustrated by
the example $L = q$, whose Euler--Lagrange equation reads $0 = 1$.
Consequently, the Dirac procedure may be employed to {\sl exhaustively determine} which actions within a
given class are consistent.  This technique is important in the constructive approach to the TSA.

A theory with constraints ${\cal C}_{\Gamma}$ initially described by a `na\"{\i}ve' Hamiltonian
$\mbox{\sffamily H\normalfont}(q_{\Delta}, p^{\Delta})$ could just as well be
described by a total Hamiltonian
\be
\mbox{\sffamily H\normalfont}_{\mbox{\scriptsize total  \normalsize}} =
\mbox{\sffamily H\normalfont} + N_{\Gamma} {\cal C }^{\Gamma}
\label{hamtot}
\ee
for arbitrary functions $N_{\Gamma}$.
For GR, the na\"{\i}ve Hamiltonian is just 0, while the total Hamiltonian is (\ref{ADMHam}).

\mbox{ }

\noindent\mbox{\bf{\Large Appendix B: EP-violating theory that admits a TSA formulation}}

\mbox{ }

\noindent Here I discuss the theory on which my EP counterexample is based.
I have known for some time that $(\nabla_AA^A)^2$-theory coupled to GR can be cast into a TSA
form quite easily with the tricks listed on page 16.
%
However, I also observed that this theory has no degrees of freedom in both the flat and curved cases
so I did not employ this example.  But the thought behind the above example
is not wasted, because, 1) the introduction of a mass term acts to introduce one
degree of freedom per space point (Isenberg \& Nester 1977a).
This is somewhat like what occurs in the transition from electromagnetism (which has two) to Proca
theory (which has three).
2) the formulation I used for $(\nabla_AA^A)^2$-theory coupled to GR is not spoiled by
the addition of the mass term.

Thus, consider the spacetime action
\be
\mbox{\sffamily I}[g_{AB}, A_A] = \int \textrm{d}\lambda \int \textrm{d}^3x\alpha
\left[
{\cal R}   + \mu
\left(
\nabla_A A^A\nabla_BA^B + m^2A^2
\right)
\right]
\mbox{ }
\label{NVA2}
\ee
which corresponds to a consistent and nontrivial theory (see Isenberg and Nester, 1977a).
Using the 1-form split (\ref{trivia}) and the split (\ref{Vderivproj3}), (\ref{Vderivproj4}) of
its covariant derivatives (all with respect to the spatial hypersurface) and refactorizing,
\be
\nabla_AA^A = \frac{1}{\alpha}
\left[
D_a(\alpha A^a) + \frac{A_{\perp}}{2}h^{ij}\delta_{\beta}h_{ij} + \delta_{\beta}A_{\perp}
\right] \mbox{ } .
\ee
I then set $\alpha A^a = \dot{v}$ and $A_{\perp}$ to be some $\phi$.
(This is, again, a case of flexibility of reformulation.)
Then the `matter' part of the Lagrangian takes the form
\be
\mbox{\sffamily L}[v_i, \dot{v}_i, \phi, \dot{\phi}, h_{ab}, \dot{h}_{ab}, \beta_i, \alpha]
= \mu\left(\frac{1}{\alpha^2}
\left[
D_a(\dot{v}^a) + \frac{\phi}{2}h^{ij}\delta_{\beta}h_{ij} + \delta_{\beta}\phi
\right]^2
+ \frac{m^2\dot{v}^2}{\alpha^2} - m^2\phi^2\right) \mbox{ } .
\ee
The gravity--1-form split lapse-uneliminated Lagrangian is then
$$
\mbox{\sffamily I}^{\mbox{\scriptsize ADM}}[h_{ab}, \dot{h}_{ab}, v_i, \dot{v}_i, \phi, \dot{\phi}, \beta_i, \alpha] =
$$
\be
\int\int \textrm{d}\lambda \alpha\sqrt{h}\textrm{d}^3x
\left[
\frac{        \mbox{\sffamily T}_{\mbox{\scriptsize GR}}[h_{ab}, \dot{h}_{ab} \beta_i]         }
     {         4\alpha^2        } +
\frac{\mu}{\alpha^2}
\left(
\left[
D_a(\dot{v}^a) + \frac{\phi}{2}h^{ij}\delta_{\beta}h_{ij} + \delta_{\beta}\phi
\right]^2
+ {m^2\dot{v}^2}\right) + R - \mu m^2\phi^2
\right],
\ee
to which the BSW procedure can be applied to obtain a reparametrization-invariant action:
$$
\mbox{\sffamily I}_{\mbox{\scriptsize BSW}}[h_{ab}, \dot{h}_{ab}, v_i, \dot{v}_i, \phi, \dot{\phi}, \beta_i] =
$$
\be
\int\int \textrm{d}\lambda \textrm{d}^3x\sqrt{h}
\sqrt{
\left(
R - \mu m^2\phi^2
\right)
\left[
\mbox{\sffamily T}_{\mbox{\scriptsize GR}}[h_{ab}, \dot{h}_{ab}, \beta_i] +
4\mu\left(\left[
D_a(\dot{v}^a) + \frac{\phi}{2}h^{ij}\delta_{\beta}h_{ij} + \delta_{\beta}\phi
\right]^2
+ m^2\dot{v}^2 \right)
\right]} \mbox{ } .
\ee
The equations encoded by this action then happen to be weakly unaffected by whether
$\dot{v}^a$ is replaced by $\delta_{\beta}v^a$.
Thus if one starts with TSA principles, and using the arbitrary 3-diffeomorphism frame symbol
$\&_{\dot{s}}$ in place of the hypersurface derivative symbol $\delta_{\beta}$,
one obtains the TSA action
$$
\mbox{\sffamily I}_{\mbox{\scriptsize TSA}}[h_{ab}, \dot{h}_{ab}, v_i, \dot{v}_i, \phi, \dot{\phi}, \dot{s}_i] =
$$
\be
\int\int \textrm{d}\lambda \textrm{d}^3x\sqrt{h}
\sqrt{
\left(
R - \mu m^2\phi^2
\right)
\left[
\mbox{\sffamily T}_{\mbox{\scriptsize GR}}[h_{ab}, \dot{h}_{ab}, \dot{s}_i] +
4\mu\left(\left[
D_a(\&_{\dot{s}}v^a) + \frac{\phi}{2}h^{ij} \&_{\dot{s}}h_{ij} + \&_{\dot{s}}\phi
\right]^2
+ m^2(\&_{\dot{s}}v)^2 \right)
\right]} \mbox{ } ,
\ee
which is consistent by the reverse of the above working.

Thus one has a consistent and nontrivial EP violating theory for geometry, a scalar and a 1-form.
It should be noted that the RWR paper missed this not on relational grounds but on simplicity
grounds: the theory has a kinetic term that is not ultralocal, has metric--matter cross-terms
and field dependence.
These properties are largely responsible for the theory having further undesirable features
(Isenberg \& Nester 1977a) physically-bizarre singularities in finite-curvature regions and
a bad flat-space limit (thus compromising the guarantee of the emergence of locally SR physics).
Thus it would be fortunate if one had a theoretical framework that excluded this;
the TSA's relationalist content, however, is unsuccessful in this respect.

\mbox{ }


\noindent{\Large{\bf References}}

\mbox{ }

\footnotesize

\noindent Alexander, H.G. (ed.) (1956). {\it The Leibniz--Clark
correspondence}. Manchester: Manchester University Press.

\noindent
Anderson, E. (2003).
Variations on the seventh route to relativity.
{\it Physical Review, D68}, 104001.

\noindent
Anderson, E. (2004a).
Strong-coupled relativity without relativity.
{\it General Relativity and Gravitation, 36}, 255-276.

\noindent
Anderson, E. (2004b).
Geometrodynamics: spacetime or space?
(Ph.D. Thesis, University of London), gr-qc/0409123.

\noindent
Anderson, E. (2004c).
The Campbell--Magaard theorem is inadequate and inappropriate
as a protective theorem for relativistic field equations,
gr-qc/0409122.

\noindent
Anderson, E. (2005).
Leibniz--Mach foundations for GR and fundamental physics.
In A. Reimer (ed.),
{\it General relativity research trends.
Horizons in world physics, Vol. 249}.
New York: Nova.

\noindent
Anderson, E, preprint. 
Relational particle models as toy models for quantum gravity and quantum cosmology, 
Submitted to {\it Proceedings of the Albert Einstein Century International Conference, 
Paris, France, July 18-22, 2005}, gr-qc/0509054.  

\noindent
Anderson, E (2006a).
Relational particle models: I. Reconciliation with standard classical and quantum theory 
{\it Classical and Quantum Gravity, 23}, 2469-2490. 

\noindent
Anderson, E, (2006b).
Relational particle models: II. Use as toy models for quantum geometrodynamics
{\it Classical and Quantum Gravity, 23}, 2491-2518

\noindent
Anderson, E. \& Barbour, J.B. (2002).
Interacting vector fields in relativity without relativity.
{\it Classical and Quantum Gravity, 19}, 3249-3262.

\noindent
Anderson, E., Barbour, J.B., Foster, B.Z., Kelleher, B. \&  \'{O} Murchadha, N. (2005).
The physical gravitational degrees of freedom.
{\it Classical and Quantum Gravity, 22}, 1795-1802.

\noindent
Anderson, E., Barbour, J.B., Foster B.Z. \& \'{O}  Murchadha, N. (2003).
Scale-invariant gravity: geometrodynamics.
{\it Classical and Quantum Gravity, 20}, 1571-1604.

\noindent
Arnowitt, R., Deser, S. \& Misner, C.W. (1962).
The dynamics of general relativity.
In L. Witten (ed.),
{\it Gravitation: an introduction to current research} (pp. 227-265).
New York: Wiley.

\noindent
Ashtekar, A. (notes prepared in collaboration with Tate,
R.) (1991). {\it Lectures on nonperturbative canonical gravity}.
Singapore: World Scientific.

\noindent
Baierlein, R.F., Sharp, D. \& Wheeler, J.A. (1962).
Three-dimensional geometry as carrier of information about time.
{\it Physical Review, 126}, 1864-1865.

\noindent
Barbour, J.B. (1986).
Leibizian time, Machian dynamics and quantum gravity.
In R. Penrose \& C.J. Isham (eds.),
{\it Quantum concepts in space and time} (pp. 236-246).
New York: Oxford University Press.

\noindent
Barbour, J.B. (1994a).
The timelessness of quantum gravity. I. The evidence from the classical theory.
{\it Classical and Quantum Gravity, 11}, 2853-2873.

\noindent
Barbour, J.B. (1994b).
The timelessness of quantum gravity. II. The appearance of dynamics in static configurations.
{\it Classical and Quantum Gravity, 11},  2875-2897.

\noindent
Barbour, J.B. (1995).
General relativity as a perfectly Machian theory.
In J.B. Barbour \& H. Pfister (eds.),
{\it Mach's principle: from Newton's bucket to quantum gravity, Vol. 6 of Einstein Studies}
(pp. 214-236).
Boston: Birkh\"{a}user.

\noindent
Barbour, J.B. (1999a).
{\it The End of Time}.
New York: Oxford University Press.

\noindent
Barbour, J.B. (1999b).
The development of Machian themes in the twentieth century.
In J. Butterfield (ed.),
{\it The Arguments of Time} (pp. 83-110).
New York: Oxford University Press.

\noindent Barbour, J.B. (personal communications to Giulini, D. and Anderson, E. in 2001--2004).

\noindent
Barbour, J.B. (2003).
Scale-invariant gravity: particle dynamics.
{\it Classical and Quantum Gravity 20}, 1543-1570,
gr-qc/0211021.

\noindent
Barbour, J.B. \& Bertotti, B. (1977).
Gravity and inertia in a Machian framework.
{\it Nuovo Cimento B, 38}, 1-27.

\noindent
Barbour, J.B. \& Bertotti, B. (1982).
Mach's principle and the structure of dynamical theories.
{\it Proceedings of the Royal Society of London A, 382}, 295-306.

\noindent
Barbour, J.B, Foster, B.Z. \& \'{O} Murchadha, N. (2002a).
Relativity without relativity.
{\it Classical and  Quantum Gravity, 19}, 3217-3248.

\noindent
Barbour, J.B., Foster, B.Z. \& \'{O} Murchadha, N. (2002b).
Contained in v1 of preprint gr-qc/0012089 of the previous citation.

\noindent
Barbour, J.B \& \'{O} Murchadha, N. (1999).
Classical and quantum gravity on conformal superspace.
gr-qc/9911071.

\noindent
Bartnik, R. \& Fodor, G. (1993).
Proof of the thin sandwich conjecture.
{\it Physical Review D, 48}, 3596-3569.

\noindent
Baumgarte, T.W. \& Shapiro, S.L. (2003).
Numerical relativity and compact binaries.
{\it Physics Reports, 376}, 41-131,
gr-qc/0211028.

\noindent
Belasco, E.P. \& Ohanian, H.C. (1969).
Initial conditions in general relativity: lapse and shift formulation
{\it Journal of Mathematical Physics, 10},  1503-1507.

\noindent
Berkeley, Bishop G. (1710)
\it The Principles of Human  Knowledge \normalfont.

\noindent
Berkeley, Bishop G.  (1721)
\it Concerning Motion (De Motu) \normalfont.

\noindent
Brans, C. \& Dicke, R. (1961).
Mach's principle and a relativistic theory of gravitation.
{\it Physical Review, 124}, 925-935.

\noindent
Butterfield, J.N. (2002)
Critical Notice.
{\it The British Journal for Philosophy of Science, 53}, 289-330, gr-qc/0103055.

\noindent
Carlip, S.J. (1998).
{\it Quantum Gravity in 2 + 1 Dimensions}
Cambridge: Cambridge University Press.

\noindent
Carlip, S.J. (2001).
Quantum gravity: a progress report.
{\it Reports on progress in physics, 64}, 885-942.

\noindent
Cartan, \'{E}. (1922).
Sur les \'{e}quations de la gravitation de Einstein.
{\it Journal de Math\'{e}matiques pures et appliqu\'{e}es, 1}, 141-203.

\noindent
Cartan, \'{E}. (1925).
La g\'{e}ometrie des espaces de Riemann,
{\it M\'{e}morial des Sciences Math\'{e}matiques, Fasc. 9}.

\noindent
Choquet-Bruhat, Y., Isenberg, J. \& York, J.W. (2000).
Einstein constraints on asymptotically Euclidean manifolds.
{\it Physical Review D, 61}, 084034.

\noindent
DeWitt, B.S. (1967).
Quantum theory of gravity, I. The canonical theory.
{\it Physical Review, 160}, 1113-1148.

\noindent
d'Inverno, R.A. \& Stachel, J. (1978).
Conformal two-structure as the gravitational degrees of freedom in general relativity.
{\it Journal of Mathematical Physics, 19}, 2447-2460.

\noindent
Dirac, P.A.M., (1951)
The Hamiltonian form of field dynamics.
{\it Canadian Journal of Mathematics ,3}, 1-23.

\noindent
Dirac, P.A.M. (1958).
The theory of gravitation in Hamiltonian form.
{\it Proceedings of the Royal Society of London A, 246}, 333-343.

\noindent
Dirac, P.A.M. (1964).
{\it Lectures on Quantum Mechanics}.
New York: Yeshiva University.

\noindent
Ehlers, J. (1987).
Folklore in relativity and what is really known.
In M.A.H. MacCallum (ed.),
{\it Proceedings, XI International Conference on GRG}  (pp. 61-71).
Cambridge: Cambridge University Press.

\noindent
Ehlers, J. \& Geroch, R.P. (2004).
Equation of motion of small bodies in relativity.
{\it Annals of Physics, 309}, 232-236.

\noindent
Ehlers, J., Pirani, F.A.E. \& Schild, A. (1972).
The geometry of free-fall and light propagation.
In L. O'Raifeartaigh (ed.),
{\it General Relativity (Synge Festschrift)} (pp. 63-84).
New York: Oxford University Press.

\noindent Einstein, A. (1915). The field equations of gravitation.
{\it Sitzungsberichte der Preussischen Akademie der Wissenschaften
}

\noindent
{\it(Physikalisch-Mathematische klasse)},
844-847.

\noindent
Einstein, A. (1916).
The foundation of the general theory of relativity.
{\it Annalen der Physik, 49}, 769-822.
%

\noindent
Einstein, A. (1918).
Dialog \"{u}ber Einw\"{a}nde gegen die Relativit\"{a}tstheorie.
{\it Die Naturwissenschaften, 6}, 697-702.

\noindent
Einstein, A. (1933).
Mein Weltbild
Amsterdam: Querido Verlag.

\noindent
Einstein, A. (1950).
{\it The Meaning of Relativity}
Princeton: Princeton University Press.

\noindent
Fierz, M. \& Pauli, W. (1939).
Relativistic wave equations for particles of arbitrary spin in an electromagnetic field.
{\it Proceedings of the Royal Society of London A, 173}, 211-232.

\noindent
Francisco, G. \& Pilati, M. (1985).
Strong-coupling quantum gravity. III. Quasiclassical approximation
{\it Physical Review D, 31}, 241-250.

\noindent
Friedrich, H. \& Rendall, A.D. (2000). The Cauchy
problem for the Einstein equations. {\it Lecture Notes in Physics,
540},  127-224, and references therein.

\noindent
Geroch, R.P. (1967).
Topology change in general relativity.
{\it Journal of Mathematical Physics, 8}, 782-786.

\noindent
Giulini, D. (1999).
The generalized thin-sandwich problem and its local solvability.
{\it Journal of Mathematical Physics, 40}, 2470-2482.

\noindent
Giulini, D. (personal communication to Barbour, J.B. in 2001).

\noindent
Green, M.B., Schwartz, J.H. \& Witten, E. (1987).
{\it Superstring theory}.
Cambridge: Cambridge University Press.

\noindent
Greene, B. Plenary talk given at Einstein Centenary Conference, Paris, 2005.

\noindent Hartle, J.B. (1995). Spacetime quantum mechanics and the
quantum mechanics of spacetime. In B. Julia \& J. Zinn-Justin
(eds.), {\it Gravitation and quantizations: Proceedings of the
1992 Les Houches Summer School}.
Amsterdam: North Holland.

\noindent Hawking, S.W. (1984). Lectures in quantum cosmology. In
B.S. DeWitt \& R. Stora (eds.) {\it Relativity, groups and
topology II} (pp. 333-379). Amsterdam: North Holland.


\noindent
Hawking, S.W. (1992).
The chronology protection conjecture.
{\it Physical Review, D46}, 603-611.

\noindent
Hawking, S.W. and Page, D.N. (1986).
Operator ordering and the flatness of the universe.
{\it Nuclear Physics B, 264}, 185-196.

\noindent
Hawking, S.W. and Page, D.N. (1988).
How probable is inflation?
{\it Nuclear Physics B, 298}, 789-809.

\noindent
Henneaux, M. (1979).
Zero Hamiltonian signature spacetimes.
{\it Bulletin de la soci\'{e}t\'{e} math\'{e}matique de Belgique, 31}, 47-63.

\noindent
Hojman, S.A. \& Kucha\v{r} K.V. (1972).
The Einstein--Hamilton--Jacobi equation and first principles.
{\it Bulletin of the American Physical Society, 17}, 450-451.

\noindent
Hojman, S.A., Kucha\v{r}, K.V. \& Teitelboim, C. (1973).
New approach to general relativity.
{\it Nature Physical Science, 245}, \normalfont 97-98.

\noindent
Hojman, S.A., Kucha\v{r}, K.V. \& Teitelboim, C. (1976).
Geometrodynamics regained.
{\it Annals of Physics,  96}, 88-135.

\noindent
Isenberg, J.A. (1981).
Wheeler-Einstein-Mach space-times.
{\it Physical Review D, 24}, 251-256.

\noindent
Isenberg, J.A \& Nester, J.M. (1977a).
The effect of gravitational interaction on classical fields: a Hamilton--Dirac analysis.
{\it Annals of Physics, 107}, 56-81.

\noindent
Isenberg, J.A. \& Nester, J.M. (1977b).
Extension of the York field equation decomposition to general gravitationally coupled fields.
{\sl Annals of Physics,  108}, 368-386.

\noindent
Isenberg, J.A., \'{O} Murchadha, N. \& York, J.W. (1976).
Initial value problem of general relativity. 3. Coupled fields and the scalar-tensor theory.
{\it Physical Review D, 13}, 1532-1537.

\noindent
Isenberg, J. \& Wheeler J.A.  (1979).
Inertia here is fixed by mass-energy there in every W-model universe.
In M. Pantaleo \& F. deFinis (eds.),
{\it Relativity, Quanta  and Cosmology in the Development of the Scientific
Thought of Albert Einstein, Vol. I} (pp. 267-293).
New York: Johnson Reprint Corporation.

\noindent
Isham, C.J. (1976).
Some quantum field theory aspects of the superspace quantization of general relativity.
{\it Proceedings of the Royal Society of London A, 351},  209-232.

\noindent
Isham, C.J. (1993).
Canonical quantum gravity and the problem of time.
In L.A. Ibort \& M.A. Rodr\'{\i}guez (eds.),
{\it Integrable systems, quantum groups and  quantum field theories} (pp. 157-288).
Dordrecht: Kluwer.

\noindent
Butterfield, J. \& Isham, C.J. (2000).
Space-time and the philosophical challenge of quantum gravity.
In C. Callender \& N. Huggett (eds.),
{\it Physics meets philosophy at the Planck scale}  (pp. 33-89).
Cambridge: Cambridge University Press.

\noindent
Jammer, M. (1993).
{\it Concepts of space.  The history of theories of space in physics.  Third, enlarged edition.}
New York: Dover.

\noindent
Kelleher, B. (2003).
Gravity on Conformal Superspace.
(Ph.D.  Thesis, University of Cork),
gr-qc/0311034.

\noindent
Kelleher, B. (2004).
Scale-invariance in gravity and implications for the cosmological constant.
{\it Classical and Quantum Gravity, 21}, 2623-39.

\noindent
Kiefer, C. (2004).
{\it Quantum Gravity}.
Oxford: Clarendon.

\noindent
Kucha\v{r}, K.V. (1972).
A bubble-time canonical formalism for geometrodynamics
{\it Journal of Mathematical Physics, 13}, 768-781.

\noindent
Kucha\v{r}, K.V. (1973).
Canonical quantization of gravity.
In W. Israel (ed.),
\it Relativity, Astrophysics and  Cosmology \normalfont (pp. 237-283).
Dordrecht: Reidel.

\noindent
Kucha\v{r}, K.V. (1974).
Geometrodynamics regained: a Lagrangian approach.
{\it Journal of Mathematical Physics, 15}, 708-715.

\noindent
Kucha\v{r}, K.V. (1976a).
Geometry of hyperspace. I.
{\it Journal of Mathematical Physics, 17}, 777-791.

\noindent
Kucha\v{r}, K.V. (1976b).
Kinematics of tensor fields in hyperspace. II.
{\it Journal of Mathematical Physics, 17}, 792-800.

\noindent
Kucha\v{r}, K.V. (1976c).
Dynamics of tensor fields in hyperspace. III.
{\it Journal of Mathematical Physics, 17}, 801-820.

\noindent
Kucha\v{r}, K.V. (1977).
Geometrodynamics with tensor sources. IV.
{\it Journal of Mathematical Physics, 18}, 1589-1597.

\noindent
Kucha\v{r}, K.V. (1981).
Canonical methods of quantization.
In C.J. Isham, R. Penrose \& D.W. Sciama (eds.),
{\it Quantum Gravity 2: A Second  Oxford Symposium}  (pp. 329-376).
Oxford: Clarendon.

\noindent
Kucha\v{r}, K.V. (1991).
The problem of time in canonical quantization of relativistic systems.
In A. Ashtekar \& J. Stachel (eds.),
{\it Conceptual problems of quantum gravity, Vol. 2 of Einstein studies} (pp. 141-171).
Boston: Birkh\"{a}user.

\noindent
Kucha\v{r}, K.V. (1992).
Time and interpretations of quantum gravity.
In G. Kunstatter, D. Vincent \& J. Williams  (eds.),
\it Proceedings of the 4th Canadian Conference on General Relativity and Relativistic Astrophysics\normalfont (pp. 211-314).
Singapore: World  Scientific.

\noindent
Kucha\v{r}, K.V. (1999).
The problem of time in quantum geometrodynamics.
In J. Butterfield (ed.),
{\it The Arguments of  Time} (pp. 169-195).
New York: Oxford University Press.

\noindent
Lachi\`{e}ze-Rey , M. and Luminet, J.P. (1995).
Cosmic Topology.
{\it Physics Reports, 254},  135-214.

\noindent
Lanczos, C. (1949).
The variational principles of mechanics.
Toronto: University of Toronto Press.

\noindent
Lichnerowicz, A. (1944).
L'int\'{e}gration des \'{e}quations de la gravitation relativiste et le probl\`{e}me des $n$ corps.
{\it Journal de }

\noindent {Math\'{e}matiques pures et appliqu\'{e}es,  23}, 37-63.

\noindent
Lovelock, D. (1971).
The Einstein tensor and its generalizations.
{\it Journal of Mathematical Physics, 12}, 498-501.

\noindent
Mach, E. (1883)
{\it Die Mechanik in ihrer Entwickelung,  Historisch-kritisch dargestellt}
Leipzig: J.A. Barth.

\noindent
Minkowski, H. (1908).
Address delivered at 80th Assembly of German Natural Scientists and Physicians, Cologne, 21 September 1908.

\noindent
Misner, C.W., Thorne, K. \& Wheeler, J.A. (1973).
{\it Gravitation}
San Francisco: Freedman.

\noindent
Misner, C.W. \& Wheeler, J.A. (1957).
Classical physics as geometry: gravitation, electromagnetism, unquantized charge, and mass as properties of curved empty space.
{\it Annals of Physics, 2},  525-603.

\noindent
Moncrief, V. \& Teitelboim, C. (1972).
Momentum constraints as integrability conditions for the Hamiltonian constraint in general relativity.
{\it Physical Review D, 6}, 966-969.

\noindent
\'{O} Murchadha, N. (2002).
Constrained Hamiltonians and local-square-root actions.
{\it International Journal of Modern Physics A, 20},  2717-2720.

\noindent
\'{O} Murchadha, N. (2003).
General relativity from the three-dimensional linear group.
gr-qc/0305038.

\noindent
Page, D.N. \& Wootters W.K. (1983).
Evolution without evolution: dynamics described by stationary observables.
{\it Physical Review D, 27}, 2885-2892.


\noindent
Penrose, R. (1979).
Singularities and time-asymmetry.
In S.W. Hawking \& W. Israel (eds.)
{\it General Relativity, an Einstein Centenary Survey}  (pp. 581-638).
Cambridge: Cambridge University Press.

\noindent
Penrose, R. \& Rindler, W. (1987).
\it Spinors and Space-time. Vol. 1. Two-spinor calculus and relativistic fields \normalfont
Cambridge: Cambridge University Press.

\noindent
Perez, A. (2003). Spin foam models for quantum gravity
{\it Classical and Quantum Gravity, 20}, R43-R104.

\noindent
Pfeiffer, H.P. \& York, J.W. (2003).
Extrinsic curvature and the Einstein constraints.
{\it Physical Review D, 67}, 044022.

\noindent
Pilati, M. (1982a).
Strong-coupling quantum gravity: an introduction.
In M. Duff \& C.J. Isham (eds.),
{\it Quantum structure of space and time} (pp. 53-69).
Cambridge: Cambridge University Press.

\noindent
Pilati, M. (1982b).
Strong-coupling quantum gravity. I. Solution in a particular gauge.
{\it Physical Review D, 26}, 2645-2663.

\noindent
Pilati, M. (1983).
Strong-coupling quantum gravity. II. Solution without gauge fixing.
{\it Physical Review D, 28}, 729-744.

\noindent
Pooley, O. (2000).
Relationalism rehabilitated? II: relativity.
http://philsci-archive.pitt.edu/archive/00000221/index.html.

\noindent
Pooley, O. (2003-4).
Comments on Sklar's ``Barbour's Relationist Metric of Time".
{\it Chronos (Proceedings of the Philosophy of Time Society)}.

\noindent
Pooley, O. \& Brown, H.R. (2002).
Relationalism rehabilitated? I: classical mechanics.
{\it British Journal for the Philosophy of Science, 53}, 183-204.

\noindent
Rainich, G.Y. (1925).
Electrodynamics in the general relativity theory
{\it Transactions of the American Mathematical Society, 27}, 106-136.

\noindent
Rindler, W. (2001).
{\it Relativity: Special, general and cosmological}
New York: Oxford University Press.

\noindent
Sakharov, A.D. (1967).
Vacuum quantum fluctuations in curved space and the theory of gravitation.
{\it Doklady Akademii Nauk S.S.S.R, 177}, 70-71.

\noindent
Savvidou, K. (2005). General relativity histories
theory. {\it Brazilian Journal of Physics, 35}, 307-315.

\noindent
Seriu, M. (1996).
Spectral representation of the spacetime structure: The ``distance'' between universes with different topologies.
{\it Physical Review D, 53}, 6902-6920.

\noindent Smolin, L. (2000). The present moment in quantum
cosmology: challenges to the arguments for the elimination of
time. In R. Durie (ed.), {\it Time and the instant} (pp. 112-143).
Manchester: Clinamen Press.

\noindent
Smolin, L. (2003),
How far are we from the quantum theory of gravity?
hep-th/0303185.

\noindent Sorkin, R.D. (2003) Causal sets: Discrete gravity (Notes
for the Valdivia Summer School). to appear in A. Gomberoff \& D.
Marolf (eds.), {\it Proceedings of the Valdivia Summer School},
gr-qc/0309009.

\noindent Stachel, J. (1989). Einstein's search for general
covariance. In D. Howard \& J. Stachel (eds.), {\it Einstein and
the history of general relativity, Vol. 1 of Einstein studies}
(pp. 63-100). Boston: Birkh\"{a}user.

\noindent Stachel, J. (1992). The Cauchy problem in general
relativity - the early years. In J. Eisenstaedt \& A.J. Kox
(eds.), {\it Historical studies in general relativity, Vol 3. of
Einstein studies} (pp. 407-418). Boston: Birkh\"{a}user.

\noindent Stewart, J.M. (1991). {\it Advanced general relativity}.
Cambridge: Cambridge University Press.

\noindent
Teitelboim, C. (1973a).
How commutators of constraints reflect the space-time structure.
{\it Annals of Physics, 79}, 542-557.

\noindent Teitelboim, C. (1973b). The Hamiltonian structure of
spacetime. (Ph.D. Thesis, Princeton).

\noindent Teitelboim, C. (1980). The Hamiltonian structure of
space-time. In A. Held (ed.), {\it General relativity and
gravitation, Vol. 1} (pp. 195-225). New York: Plenum Press.

\noindent
Teitelboim, C. (1982).
Quantum mechanics of the gravitational field.
{\it Physical Review D, 25}, 3159-3179.

\noindent
Teitelboim, C. \& Zanelli, J. (1987).
Dimensionally continued topological gravitation theory in Hamiltonian form.
{\it Classical and Quantum Gravity, 4}, L125-129.

\noindent
Unruh, W.G. \&  Wald, R.M. (1989).
Time and the interpretation of canonical quantum gravity.
{\it Physical Review D, 40}, 2598-2614.

\noindent
Vermeil, H. (1917).
Notiz \"{u}ber das mittlere Kr\"{u}mmungmass einer $n$-fach augelehnten Riemann'schen Mannigfaltigkeit.
{\it Nachrichten von der K\"{o}nigliche Gesellschaft der Wissenschaften zu G\"{o}ttingen}
(Mathematisch-Physikalische Klasse), 334-344.

\noindent Wald, R.M. (1984). \it General relativity. \normalfont
Chicago: University of Chicago Press.

\noindent
Weyl, H. (1918).
{\it Sitzungsberichte der Preussichen akademie der Wissenschaften},
465-478.

\noindent Weyl, H. (1921) \it Space-time-matter. \normalfont (4th
ed.). Berlin: Springer.

\noindent
Wheeler, J.A. (1962).
{\it Geometrodynamica}.
New York: Academic Press.

\noindent Wheeler, J.A. (1964a). Geometrodynamics and the issue of
the final state. In B.S. DeWitt \& C.M. DeWitt (eds.), {\it
Groups, relativity and Ttopology}  (pp. 317-520). New York: Gordon
and Breach.

\noindent
Wheeler, J.A. (1964b).
Gravitation as geometry.
In H.-Y. Chiu \& W.F. Hoffmann (eds.) {\it Gravitation and relativity}  (pp. 303-349).
New York: Benjamin.

\noindent
Wheeler, J.A. (1968).
Superspace and the nature of quantum geometrodynamics.
In C. DeWitt \& J.A. Wheeler (eds.),
{\it Battelle rencontres: 1967 lectures  in mathematics and physics}  (pp. 242-307).
New York: Benjamin.

\noindent
Wheeler, J.A. (1988).
Geometrodynamic steering principle reveals the determiners of inertia.
{\it International Journal of Modern Physics A, 3}, 2207-2247.

\noindent
Will, C.M. (2001).
The confrontation between general relativity and experiment.
{\it Living Reviews in Relativity, 4}.

\noindent
Winicour, J. (2001).
Characteristic evolution and matching.
{\it Living Reviews in Relativity, 3}.

\noindent
York, J.W. (1971).
Gravitational degrees of freedom and the initial value problem.
{\it Physical Review Letters, 26}, 1656-1658.

\noindent
York, J.W. (1972).
The role of conformal 3-geometry in the dynamics of gravitation.
{\it Physical Review Letters, 28}, 1082-1085.

\noindent York, J.W. (1973).
Conformally invariant decomposition of symmetric tensors on
Riemannian manifolds and the initial-value problem of general relativity.
{\it Journal of Mathematical Physics, 14}, 456-465.

\noindent
York, J.W. (1999).
Conformal `thin sandwich' data for the initial value problem.
{\it Physical Review Letters, 82}, 1350-1353.



\end{document}